\documentclass[%
preprint,
showpacs,
amsmath,amssymb,
aps,
prb,
]{revtex4-1}
\usepackage{graphicx}
\usepackage{dcolumn}
\usepackage{bm}
\usepackage{color}
\usepackage{verbatim}

\begin{document}

\preprint{APS/123-QED}

\title{Non-static effects in ordered and disordered quantum spin systems I: theoretical formulation, energy gap and non-extensive terms of ground-state energy of the ferromagnetic Ising model in a transverse field}

\author{Yang Wei Koh}
\email{patrickkyw@gmail.com}
\affiliation{Bioinformatics Institute, 30 Biopolis Street, No. 07-01, Matrix, Singapore 138671}

\date{\today}

\begin{abstract}
In the path integral formulation of the partition function of quantum spin models, most current treatments employ the so-called static approximation to simplify the process of summing over all possible paths. Although sufficient for studying the thermodynamic aspects of the system, static approximation ignores the contributions made by time-dependent, or non-static, fluctuations in the paths of the path integral. This non-static component is very small relative to the static part, and its careful treatment is necessary for the calculation of small non-extensive quantities such as the energy gap within the path integral framework. We propose a formalism for incorporating non-static effects into the path integral calculation of a class of spin models whose partition functions are reducible to the trace of a single spin (often known as the effective Hamiltonian). We first show that the time-dependent behavior of the single spin trace is governed by the Pauli equation. Time-dependent perturbation theory is used to obtain a perturbative expansion of the solution of the Pauli equation, and then for the single spin trace. This gives us a perturbative expansion of the path integral which can be integrated systematically using standard techniques. In this paper, we develop the theoretical framework outlined above in detail and apply it to a simple ordered spin model, the infinite-range ferromagnetic Ising model in a transverse field. We calculated two non-extensive quantites with this non-static approach: the $N^0$ and $N^{-1}$ terms of the ground-state energy ($N$=number of spins) and the energy gap between the ground and first-excited states. We checked our results by comparing with those of Holstein-Primakoff transform and numerical diagonalization of the Hamiltonian. The application of the method to quantum spin-glasses is briefly discussed.

\end{abstract}
\pacs{}
\maketitle


\section{Introduction}
\label{sec.intro}

The study of quantum spin systems is recently receiving much attention\cite{Perdomo-Ortiz12,Liu15,Nishimura16,Takahashi17} due to the prospects of using quantum annealing\cite{Kadowaki98,Farhi01,Das08} to perform quantum computation. The path integral formulation of the partition function of such systems is now a standard technique for studying and analyzing the variety of different quantum spin models studied in quantum annealing\cite{Jorg08,Jorg10,Bapst12,Seki12}. In this method, one is faced with the task of summing over all possible paths of an order parameter where each path evolves along an additional time dimension introduced by the Suzuki-Trotter decomposition. In their early study of the random quantum Heisenberg model, Bray and Moore\cite{Bray80} introduced the so-called static approximation as a first step in overcoming the difficulty posed by the integration of paths. In this approximation, one assumes that the paths are constant in time and neglects their time-dependence. Using the static approximation, one can easily calculate various thermodynamic properties of a system as well as determine the phase diagram, and its use is now normal procedure in the analysis of the many different ordered \cite{Seki12,Jorg10,Bapst12} and disordered \cite{Thirumalai89,Nishimori96,Jorg08,Bray80} quantum spin models.

Despite its widespread acceptance and usage, many have tried to avoid making and to go beyond the static approximation. The class of systems known as $p$-spin interaction models \cite{Jorg10,Jorg08,Dobrosavljevic90,Goldschmidt90a,DeCesare96,Obuchi07} has been widely studied because in the limit $p\rightarrow\infty$ static approximation becomes exact, and this allows one to calculate the transition into the ordered phase without worrying about complications due to non-staticity. When applied to finite-$p$ senarios or to other models, however, static approximation displays peculiarities such as non-vanishing of the entropy at zero temperature \cite{Thirumalai89} and incorrect prediction of the phase diagram \cite{Dobrosavljevic90,DeCesare96,Takahashi07}. Early attempts to improve upon the static approximation used numerical methods such as exact spin-summation or Monte-Carlo techniques \cite{Usadel87,Goldschmidt90b} to obtain the time-dependence of the order parameters. Near the critical point, analytic methods such as quantum linked-cluster expansion can also be used to calculate the decay of the parameters with time \cite{Miller93}. An efficient numerical algorithm for computing the exact time-dependence of the order parameter by solving the mean-field equation has also been developed\cite{Grempel98, Rozenberg98}. The effects of non-staticity on the phase diagram have also been examined under the framework of Landau expansion where the time-dependence of the order parameters are systematically integrated out in the path integral\cite{Takahashi07}.

In these earlier works, the focus has mainly been on the role played by non-staticity in determining the free energy and hence the thermodynamic properties of the system. However, in addition to thermodynamic quantities, which are extensive in nature, there are also some non-extensive (i.e., not scaling linearly with system size $N$) ones of interest that cannot be calculated simply by considering the free energy. One such quantity is the energy gap between the ground and the first excited-state. As the gap arises from the flipping of just a finite number of spins in the ground-state configuration, it is very small compared to the extensive free energy. In the context of quantum annealing, the minimum gap along an annealing trajectory determines the success rate of the annealing process\cite{Sei07,Ohzeki11}, and there is currently much interest in methods for calculating the gap that are applicable to large spin systems \cite{Seki12,Seoane12,Koh16}. Another example is the non-extensive part of the ground-state energy. The free energy is actually only the leading term of the ground-state energy, and to attain better accuracies for the latter one must compute non-extensive correction terms with higher powers of $N^{-1}$. Knowledge of the $N^{-1}$ term of the ground-state energy is also essential for calculating the entanglement properties and the scaling exponents of correlation functions of finite-size systems\cite{Dusuel04,Dusuel05}. These information are not obtainable from the free energy and their calculations require different treatments\cite{Dusuel04,Dusuel05,Wegner94,Glazek93,Glazek94}. 

The purpose of this paper is to incorporate non-staticity into the path integral of spins with the aim of calculating quantities which are non-extensive in nature. We shall follow the traditional perturbative approach. In the classic Feynman kernel, a path $x(t)$ is usually expanded as $x_{cl}(t)+\lambda y(t)$ where $x_{cl}(t)$ is the classical trajectory, $y(t)$ is the deviation describing quantum fluctuations, and $\lambda$ is a small expansion parameter usually taken to be $\sqrt{\hbar}$. Quantum effects such as the decay probability due to tunneling can be obtained by performing the path  integral over all possible $y(t)$. In a similar vein, we propose that the path of an order parameter be divided into a static and a non-static component (c.f. Eq. (\ref{eq.III.B.01}) below). The role of $\lambda$ is now played by the inverse system size $1/\sqrt{N}$\,\cite{note. why sqrt of N}. By systematically expanding the path integral in powers of $\lambda$ around the static portion of the path	 and then integrating out the non-static parts, non-extensive quantities such as the energy gap and corrections to the free energy can be calculated. 

An important difference between the Feynman kernel and the path integral of spins is in the way of performing the expansion about the classical or static part of the path. In the Feynman kernel, expansion about a classical trajectory gives rise to the time evolution of a small volume element in the vicinity of the trajectory\cite{Takahashi04}. This volume element is governed by Hamilton's equations of motion for the monodromy matrix \cite{Goldstein01,Brewer97}, and its effect is to contribute the so-called the van Vleck determinant\cite{Schulman96} to the classical probability amplitude. On the other hand, there is no analogous version of the monodromy matrix in the case of path integral of spins. Instead, for many of the spin models of interest, one has to evaluate the trace of a single spin with a time-dependent Hamiltonian. The first  contribution of this paper is to revisit and make a careful analysis of this single spin trace. It is found that the time-dependence of the trace can actually be interpreted as the solution of the Pauli equation for spinor\cite{Bandyopadhyay08}. The non-static component of the order parameter plays the role of an external, time-dependent magnetic field governing the temporal evolution of the Pauli spinor. Once the single spin trace has been recast in this form, the usual time-dependent perturbation theory\cite{Ballentine98} can be used to obtain a perturbative expansion for the solution of the spinor, and subsequently expansion for the single spin trace and ultimately for the path integral itself. 

The second contribution of this paper is in generalizing the traditional partition function to calculate the energy gap. The partition function is defined as 
\begin{equation}
\mathrm{Tr}
\left(
e^{-\beta H}
\right)
=
D_0
e^{-\beta E_0}
+
D_1
e^{-\beta E_1}
+
\cdots
,
\label{eq.intro.01}
\end{equation}
where `Tr' denotes taking the trace, $\beta$ is the inverse temperature, $H$ is the Hamiltonian of the system, and $E_0$, $E_1$ are the ground and first excited-state energies, respectively. $D_0$ and $D_1$ are coefficients depending on the degeneracies of the energy levels. In the limit $\beta\rightarrow\infty$, the leading term $e^{-\beta E_0}$ dominates and one obtains the well-known formula relating the ground-state energy to the partition function (c.f. Eq. (\ref{eq.II.B.01}) below). We consider generalizing Eq. (\ref{eq.intro.01}) as 
\begin{equation}
\mathrm{Tr}
\left[
f\left(A,e^{-\beta H}\right)
\right]
=
\tilde{D}_0(A)
e^{-\beta E_0}
+
\tilde{D}_1(A)
e^{-\beta E_1}
+
\cdots
,
\label{eq.intro.02}
\end{equation}
where $A$ and $f$ are, respectively, an operator and a functional form which are to be carefully chosen to suit the system under investigation. The coefficients $\tilde{D}_0$ and $\tilde{D}_1$ are now dependent on the operator $A$. The idea is to choose $A$ and $f$ in such a way that $\tilde{D}_0$ vanishes while $\tilde{D}_1$ remains non-zero and finite, making $e^{-\beta E_1}$ instead of $e^{-\beta E_0}$ the leading term. One can then repeat the usual prescription of taking the logarithm followed by the limit $\beta\rightarrow\infty$ to obtain a relation between $E_1$ and $\mathrm{Tr}[f(A,e^{-\beta H})]$\,\cite{note.concerning generalized partition function}. Once $E_1$ is known, the energy gap follows immediately by subtracting away the ground-state energy $E_0$. This strategy for formulating the gap has been used in calculating the energy splitting due to tunneling in a one-dimensional double-well potential\cite{Zinn-Justin05}. Here, we extend the basic idea and apply it with specific forms of $\mathrm{Tr}[f(A,e^{-\beta H})]$ to the case of quantum spin models.

One advantage of the approach presented in this paper is that it is applicable as long as the path integral is reducible to a form involving a single spin trace. It can hence be used for both ordered and disordered models. There are currently many methods for handling ordered systems for large system sizes, both numerically\cite{Seki12,Jorg10} and analytically\cite{Seoane12,Dusuel04,Dusuel05}. On the other hand, as the Hamiltonians of disordered systems do not commute with the total angular momentum operator, one must diagonalize the full Hamiltonian matrix whose dimension scales exponentially with the number of spins, thereby limiting numerical studies to small system sizes\cite{Jorg08,Hogg03,Takahashi10a,Takahashi10b}. It is also not immediately apparent how to generalize analytical methods such as the Holstein-Primakoff transformation\cite{Seoane12,Dusuel04,Dusuel05} or the continuous unitary transformation \cite{Dusuel04,Dusuel05} to disordered systems. Currently, quantum Monte-Carlo simulation is usually used to study generic disordered spin systems for large system sizes\cite{Young08,Young10}. We think that our approach here can contribute towards the analysis of certain classes of such systems. 

This paper is the first of a two-part work on the effects of non-staticity in quantum spin models based on the ideas outlined above.  In this paper, the focus is to establish the theoretical framework of our proposed method and then illustrate its application using a simple model: the infinite-range ferromagnetic Ising model in a transverse field. The task is to calculate the energy gap and the corrections to the free energy of this model. For this simple model, the results that we derive here with our method can also be obtained using other approaches, thus allowing us to check the correctness of our results. The application of the method to a disordered model will be presented in a second paper.

The rest of the paper is organized as follows. In Sec. \ref{sec.background}, we define our model, give a brief review of the path integral of quantum spin systems, and then summarize the results of static approximation. Sec. \ref{sec.nonstaticity} focuses on theory and constitutes the heart of the paper. It is divided into three parts. We first formulate the time-dependence of the single spin trace in terms of the solution of an ordinary differential equation, the Pauli equation. We then introduce our non-static ansatz, and give a detailed presentation of time-dependent perturbation theory.  In the third part, the perturbation theory is used to solve the Pauli equation and obtain the perturbative expansion of the single spin trace. This expansion of the trace contains our central result and is used extensively in the remaining parts of the paper. In Sec. \ref{sec.E0}, we calculate the non-extensive terms of the ground-state energy using the expansion. Sections \ref{sec.gap 1} and \ref{sec.gap 2} each presents a  different way of calculating the energy gap based on two different generalized partition functions. After deriving the formulae relating the first excited-state energy to the proposed functions, the latter are formulated as a path integrals and then evaluated perturbatively in the same way as in Sec. \ref{sec.E0} for the normal partition function. Sec. \ref{sec.discussions} discusses and concludes the paper.

\section{Background and motivation}
\label{sec.background}

\subsection{Ferromagnetic model and Holstein-Primakoff approach}
\label{sec.sub.model}
The Hamiltonian of the infinite-range ferromagnetic Ising model in transverse field is given by
\begin{equation}
H=-\frac{J}{N}\left(\sum_{i=1}^{N}\sigma_i^{z}\right)^2 -\Gamma \sum_{i=1}^{N}\sigma_i^x,
\label{eq.II.A.01}
\end{equation}
where $\sigma_i^{\alpha}$ $(\alpha=x,y,z)$ is the $\alpha$-direction Pauli matrix of the $i$th spin, $N$ is the total number of spins, and $J$ and $\Gamma$ are, respectively, the strengths of the ferromagnetic coupling and transverse field. The first term of $H$ describes a set of $N$ spins all interacting identically between each other in the $z$-direction, while the second term introduces a non-commuting, external field along the $x$-direction.

It is difficult to analytically compute the exact $E_0$ when $N$ is large but finite. Several approximation schemes such as mean-field theory \cite{Das06}, variational-semiclassical approach \cite{Dusuel05}, and static approximation \cite{Jorg10,Bapst12,Seki12} exist, all of which allow one to obtain only the extensive (i.e., linear in $N$) part of $E_0$. In the following, we briefly discuss the Holstein-Primakoff (HP) transformation approach \cite{Dusuel04,Dusuel05,Das06}. The HP transform is interesting because it captures certain quantum effects, such as the non-extensive part of the ground-state energy and the excitation gap, which are not accounted for by the other methods. In this approach, one defines bosonic operators $b$ and $b^{\dagger}$ such that
\begin{equation}
S^z + i S^y = \sqrt{s-n}\,\,b,\,\,\,\,\, S^z - i S^y =b^{\dagger}\,\, \sqrt{s-n},\,\,\,\,\,S^x =s-n,
\label{eq.II.A.02}
\end{equation}
where $S^{\alpha}=\sum_{i}\sigma_i^{\alpha}$, $n=b^{\dagger}b$, and $s$ is the angular momentum quantum number. The operators $b$ and $b^{\dagger}$ satisfy $[b,b^{\dagger}]=1$. The transformation Eq. (\ref{eq.II.A.02}) is useful when one considers a state whose quantum fluctuations are small compared to its macroscopic component. For the ground-state of $H$, $s\propto N$, and assuming that the number of excitations is much smaller than $N$ (i.e., $\langle b^{\dagger}b \rangle \ll N$), one can approximate $\sqrt{s-n}\approx \sqrt{s}$ in Eq. (\ref{eq.II.A.02}). Substituting Eq. (\ref{eq.II.A.02}) into Eq. (\ref{eq.II.A.01}), the Hamiltonian $H$ (expressed in terms of $b$ and $b^{\dagger}$) is then expanded in powers of $N$. The leading part of the Hamiltonian (comprising of terms of order $N^1$ and $N^0$) is finally diagonalized by a Bogoliubov transformation using new operators $\gamma$ and $\gamma^{\dagger}$, giving
\begin{equation}
H=
\left\{
\begin{array}{lcc}
 -N \Gamma + \sqrt{\Gamma(\Gamma-2J)} - \Gamma +2\sqrt{\Gamma(\Gamma-2J)}\,\, \gamma^{\dagger}\gamma + O(N^{-1}) & \mathrm{if} & \Gamma \ge2J, \\
 -N \frac{(2J)^2+(\Gamma)^2}{4J} + \sqrt{(2J)^2-(\Gamma)^2} - 2J  +2\sqrt{(2J)^2-(\Gamma)^2}\,\, \gamma^{\dagger}\gamma + O(N^{-1}) & \mathrm{if} & \Gamma <2J, \\
\end{array}
\right.
\label{eq.II.A.03}
\end{equation}
where $\Gamma\ge 2J$ ($\Gamma< 2J$) is the paramagnetic (ferromagnetic) phase. In Eq. (\ref{eq.II.A.03}), the first term proportional to $N$ is the extensive part of the ground-state energy, also obtainable by the various other methods mentioned above. Note, however, that one obtained also the $N^0$ terms, giving the leading correction to the extensive energy and the excitation gap (the coefficient of $\gamma^{\dagger}\gamma$). These two results are not obtainable from, say, mean-field theory. 

The HP approach has its limitations. The transformation Eq. (\ref{eq.II.A.02}) cannot be applied to, for instance, disordered systems such as the Sherrington-Kirkpatrick model where the spin-spin interactions are not the same between every single pair of spins. Furthermore, as pointed out by Dusuel and Vidal\cite{Dusuel05}, it seems difficult to go beyond the $N^0$ term using this framework because the Hamiltonian is no longer quadratic starting from order $N^{-1}$ and hence cannot be diagonalized by a Bogoliubov transformation.

\subsection{Path integral representation of partition function}
\label{sec.sub.path integral representation}

Another way to compute $E_0$ is via the relation,
\begin{equation}
E_0=\lim_{\beta\rightarrow \infty}-\frac{1}{\beta}\ln Z,
\label{eq.II.B.01}
\end{equation}
where $\beta$ is the inverse temperature and $Z$ is the partition function given by,
\begin{equation}
Z=\mathrm{Tr}\left(e^{-\beta H}\right),
\label{eq.II.B.02}
\end{equation}
where `Tr' denotes taking the trace of the operator $e^{-\beta H}$. The partition function $Z$ can be evaluated using standard path integral techniques routinely used in the treatment of quantum spin systems\cite{Jorg10,Bapst12,Seki12}. Applying the Suzuki-Trotter decomposition, one has,
\begin{eqnarray}
Z&=&\lim_{M\rightarrow \infty} Z_M \nonumber\\
 &=& \lim_{M\rightarrow \infty} \mathrm{Tr}
\left(
\left[
e^{\frac{\beta J}{M N } \left(\sum_i \sigma_i^z\right)^2}
e^{\frac{\beta \Gamma}{M}\sum_i \sigma_i^x}
\right]^{M}
\right).
\label{eq.II.B.03}
\end{eqnarray}
Resolutions of identity in the $z$-basis are inserted between each pair of exponentials, allowing the Pauli matrices $\sigma_i^z$ in $\left(\sum_i \sigma_i^z\right)^2$ to be evaluated in terms of Ising variables. The resulting quadratic terms are then linearized by Hubbard-Stratonovich transformations, giving,

\begin{equation}
Z_M=
\left(\sqrt{\frac{\beta J N}{\pi M}}\right)^{M}
\prod_{\kappa=0}^{M-1}
\int_{-\infty}^{\infty} dm_{\kappa}
\exp\left(-\frac{\beta J N}{M}\sum_{\kappa=0}^{M-1}m_{\kappa}^2\right)
\left(
\sum_{\sigma=\pm1}
\langle\sigma
|
\left[
\prod_{\kappa=0}^{M-1}
e^{\frac{1}{M}(\beta \Gamma \sigma^x + 2\beta J m_{\kappa} \sigma^z)}
\right]
|
\sigma\rangle
\right)^N
,
\label{eq.II.B.04}
\end{equation}
where  $m_{\kappa}$ is the order parameter (magnetization) introduced by the linearization at the $\kappa$th Trotter slice, $\sigma$ is an Ising variable taking values $\pm1$, and $|\sigma\rangle$ is the eigenvector of $\sigma^z$ corresponding to the eigenvalue $\sigma$. In arriving at the second term in the integrand of Eq. (\ref{eq.II.B.04}), we have `backtracked' by reinstating the operator $\sigma^z$ and by withdrawing all the resolutions of identity.

From Eqs. (\ref{eq.II.B.03}) and (\ref{eq.II.B.04}), we see that in the limit $M\rightarrow\infty$, the partition function $Z$ takes the form of a path integral where the sum is over all possible trajectories of $m_{\kappa}$ along the $\kappa$, or time, dimension. Eq. (\ref{eq.II.B.04}) is an exact relation. If the sum over all possible paths of $m_{\kappa}$ is performed exactly, the exact $Z$ is obtained. In practice, however, one resorts to approximations. Eq. (\ref{eq.II.B.04}) serves as the starting point for our consideration of non-static effects.  

The second term in the integrand of Eq. (\ref{eq.II.B.04}) takes a simple form involving the variables of just a single spin. An $N$-body problem has therefore been reduced to a single body one. This property, however, is not particular to the simple ferromagnetic model which we have chosen to consider here. A wide range of models, such as those with  frustrated couplings (e.g., disordered models) or with different lattice topolgy (e.g., Bethe lattice), can also be reduced in a similar manner. Hence, the formalism which we will be developing is applicable whenever one can bring the partition function to a form analogous to Eq. (\ref{eq.II.B.04}) involving a single spin.

In the existing literature, the single spin term is usually developed one step further. The operators $\sigma^z$ and $\sigma^x$ are explicitly evaluated in terms of Ising variables, resulting in a classical one-dimensional Ising model with uniform nearest-neighbor coupling and $m_{\kappa}$-dependent external field at the $\kappa$th spin site. We shall, however, consider the form given in Eq. (\ref{eq.II.B.04}) as it offers a slightly different perspective and opens up another approach for evaluating this term.

\subsection{Static approximation}
\label{sec.sub.static}

We conclude this section with a discussion of the static approximation. Essentially, one neglects all paths except those where the magnetization remains constant (or static) throughout the entire time interval, i.e.,
\begin{equation}
m_{\kappa} 
\stackrel{\mathrm{s.a.}}{\longrightarrow}
m_s, \hspace{1cm} \kappa=0,\dots,M-1,
\label{eq.II.C.01}
\end{equation}
where $\stackrel{\mathrm{s.a.}}{\longrightarrow}$ denotes static approximation, and $m_s$ denotes the static magnetization. With the ansatz Eq. (\ref{eq.II.C.01}), the single spin trace in Eq. (\ref{eq.II.B.04}) becomes trival,
\begin{equation}
\sum_{\sigma=\pm1}
\langle \sigma |
\left[
\prod_{\kappa=0}^{M-1}
e^{\frac{1}{M} ( \beta \Gamma \sigma^x + 2\beta Jm_{\kappa} \sigma^z  )  }
\right]
| \sigma \rangle
\stackrel{\mathrm{s.a.}}{\longrightarrow}
\sum_{\sigma=\pm1}
\langle \sigma |
e^{\beta \Gamma \sigma^x + 2\beta Jm_{s} \sigma^z  }
| \sigma \rangle
=
2\cosh\sqrt{(\beta\Gamma)^2 + (2\beta J m_s)^2}
.
\label{eq.II.C.02}
\end{equation}
Without Eq. (\ref{eq.II.C.01}), the exponents on the left hand side of Eq. (\ref{eq.II.C.02}) do not commute and cannot be combined into a single exponent. The partition function becomes,
\begin{equation} 
Z
\stackrel{\mathrm{s.a.}}{\longrightarrow}
\mathrm{const.}
\times
\int dm_s
\exp
\left(
-\beta N f_s
\right),
\label{eq.II.C.03}
\end{equation} 
where one integrates over all possible static paths, and,
\begin{equation} 
f_s=Jm_s^2 -\frac{1}{\beta} \ln 2 \cosh \sqrt{ (\beta\Gamma)^2 + (2\beta J m_s)^2  },
\label{eq.II.C.04}
\end{equation} 
is the static free energy per spin. In the limit $N\rightarrow \infty$, Eq. (\ref{eq.II.C.03}) is evaluated using the method of steepest descent. The stationary condition $\partial f_s/\partial m_s=0$ gives, 
\begin{equation} 
m_s
\left(
1
-
\frac{2J \tanh \sqrt{ (\beta\Gamma)^2 + (2\beta J m_s)^2  }    }{ \sqrt{ \Gamma^2 + (2 J m_s)^2  } }
\right)
=
0.
\label{eq.II.C.05}
\end{equation} 
Solution of Eq. (\ref{eq.II.C.05}) gives the static path with the greatest contribution to the integral of Eq. (\ref{eq.II.C.03}). From Eq. (\ref{eq.II.B.01}), we are interested in the limit $\beta\rightarrow \infty$. The solution is then,
\begin{equation}
m_s
=
\left\{
\begin{array}{ccc}
0 & \mathrm{for} & \Gamma \ge2J, \\
\pm\sqrt{1-\left(\frac{\Gamma}{2J}\right)^2} & \mathrm{for} & \Gamma<2J, \\
\end{array}
\right. 
\label{eq.II.C.06}
\end{equation}
and the ground-state energy is,
\begin{equation}
E_0
\stackrel{\mathrm{s.a.}}{\longrightarrow}
N f_s
=
\left\{
\begin{array}{ccc}
-N \Gamma  & \mathrm{for} & \Gamma \ge2J, \\
-N \frac{(2J)^2+\Gamma^2}{4J}  & \mathrm{for} & \Gamma <2J. \\
\end{array}
\right.
\label{eq.II.C.07}
\end{equation}
Comparing with the results of HP transform, we see that static approximation gives the extensive term of $E_0$, but not the order $N^0$ correction term.

\section{Non-staticity: Theoretical formulation}
\label{sec.nonstaticity}

\subsection{Spinor dynamics in a time-dependent, external field}
\label{sec.sub.ode interpretation}

The starting point of our consideration is the single spin trace $\mathcal{T}$,
\begin{equation}
\mathcal{T}
=
\sum_{\sigma=\pm1}
\langle \sigma |
\left[
\prod_{\kappa=0}^{M-1}
e^{\frac{1}{M} ( \beta \Gamma \sigma^x + 2\beta Jm_{\kappa} \sigma^z  )  }
\right]
| \sigma \rangle.
\label{eq.III.A.01}
\end{equation}
To motivate our discussion, let us first consider an ordinary differential equation,
\begin{equation}
\frac{d\textbf{v}(t)}{dt}=\textbf{F}(t)\textbf{v}(t),
\label{eq.III.A.02}
\end{equation}
where $\textbf{v}(t)$ is a $d$-dimensional column vector at the time $t$, and $\textbf{F}(t)$ is a $d\times d$ matrix independent of $\textbf{v}(t)$ but possibly dependent on $t$. To advance $\textbf{v}(t)$ by a small time step $\Delta t$ under the equation of motion Eq. (\ref{eq.III.A.02}), we have, 
\begin{eqnarray}
\textbf{v}(t+\Delta t)
&=&
\left[
\textbf{I}
+
\textbf{F}(t)
\Delta t
\right]
\textbf{v}(t) + O[(\Delta t)^2]
\nonumber \\
&\approx&
e^{
\textbf{F}(t)
\Delta t
}
\textbf{v}(t),
\label{eq.III.A.03}
\end{eqnarray}
where $\textbf{I}$ is the identity matrix. The solution of $\textbf{v}(t)$ at a later time $t+T$ is obtained by repeated application of Eq. (\ref{eq.III.A.03}),
\begin{equation}
\textbf{v}(t+T)
=
\prod_{\kappa=0}^{M-1}
e^{\mathbf{F}(t+\kappa \Delta t)\Delta t}
\textbf{v}(t),
\label{eq.III.A.04}
\end{equation}
where $T=M\Delta t$, and $O[(\Delta t)^2]$ terms can be ignored in the limit $\Delta t\rightarrow0$. The matrix product sequence $\prod_{\kappa} e^{\mathbf{F}(t+\kappa \Delta t)\Delta t} $ is known as the fundamental matrix of the ordinary differential equation Eq. (\ref{eq.III.A.02}). The fundamental matrix propagates an initial condition $\mathbf{v}(t)$ to a later time $t+T$. 

Returning to $\mathcal{T}$, we see that the product sequence in Eq. (\ref{eq.III.A.01}) is none other than the fundamental matrix of the differential equation,
\begin{equation}
\frac{d|\psi(t)\rangle}{dt}
=
\mathcal{H}(t)
|\psi(t)\rangle,
\label{eq.III.A.05}
\end{equation}
where $|\psi(t)\rangle$ is a two-dimensional Pauli spinor at time $t$, and 
\begin{equation}
\mathcal{H}(t)
=
\beta\Gamma\sigma^x
+
2 \beta J m(t) \sigma^z,
\label{eq.III.A.06}
\end{equation}
where $m(t)$ is the magnetization at time $t$. If one interprets $\mathcal{H}(t)$ as a Hamiltonian, then Eq. (\ref{eq.III.A.05}) has the form of the Pauli equation describing the evolution of a spinor under a time-dependent, external field $m(t)$. The trace $\mathcal{T}$ is therefore the sum of the eigenvalues of the fundamental matrix of Eq. (\ref{eq.III.A.05}) between time $t=0$ and 1. 

We divide the calculation of $\mathcal{T}$ into two steps. First, one specifies a basis and solve for the trajectories of each of the basis vectors under the equation of motion Eq. (\ref{eq.III.A.05}). In the basis of $\sigma^z$, for instance, one calculates $|\sigma(t)\rangle$, the solution at time $t$ of the $\sigma$-eigenvector of $\sigma^z$, subjected to the initial condition $|\sigma(0)\rangle=|\sigma\rangle$. Second, the autocorrelation of each of the basis vector is computed at $t=1$, and then summed, i.e.,
\begin{equation}
\mathcal{T}
=
\sum_{\sigma=\pm1}
\langle
\sigma(0)
|
\sigma(1)
\rangle.
\label{eq.III.A.07}
\end{equation}

The quantity $m(t)$ plays a dual role. On one hand, $m(t)$ is an integration variable in the path integral Eq. (\ref{eq.II.B.04}), playing the role of a path. On the other hand, in the Pauli equation Eq. (\ref{eq.III.A.05}), $m(t)$ plays the role of an external field in the time evolution of the spinor. When calculating $\mathcal{T}$ using Eq. (\ref{eq.III.A.07}), $m(t)$ remains fixed; when summing over paths, one calculates a $\mathcal{T}$ for each path $m(t)$.

\subsection{Non-static ansatz and time-dependent perturbation theory}
\label{sec.sub.ansatz}

We now solve the Pauli equation Eq. (\ref{eq.III.A.05}) using perturbation theory. We propose the non-static ansatz for the path,
\begin{equation}
m(t)=m_s+\lambda m_d(t),
\label{eq.III.B.01}
\end{equation}
where $m_s$ is the static part and $m_d(t)$ is the non-static part of $m(t)$. $\lambda$ is a small parameter (later shown to be $1/\sqrt{N}$). Eq. (\ref{eq.III.B.01}) means that the non-static part acts as a perturbation to the static part. $\mathcal{H}(t)$ can then be written as, 
\begin{equation}
\mathcal{H}(t)
=
\mathcal{H}_s
+
\lambda
\mathcal{H}_d(t),
\label{eq.III.B.02}
\end{equation}
where $\mathcal{H}_s=\beta \Gamma \sigma^x + 2 \beta J m_s \sigma^z$ and $\mathcal{H}_d(t)=2\beta J m_d(t) \sigma^z$. Hamiltonians of the form Eq. (\ref{eq.III.B.02}) where the time-independent part is perturbed by a small time-dependent term can be treated using time-dependent perturbation theory \cite{Ballentine98}.

Let $\varepsilon_n$ be an eigenvalue of $\mathcal{H}_s$ and $|n\rangle$ be the corresponding eigenvector. As the set $\{|n\rangle\}$ forms a complete basis, expand $|\psi(t)\rangle$ of Eq. (\ref{eq.III.A.05}) as,
\begin{equation}
|\psi(t)\rangle
=
\sum_n
\psi_n(t)e^{\varepsilon_n t}|n\rangle.
\label{eq.III.B.03}
\end{equation}
In Eq. (\ref{eq.III.B.03}), $e^{\varepsilon_n t}$ takes care of the time-dependence due to $\mathcal{H}_s$ while $\psi_n(t)$ takes care of that due to $\lambda\mathcal{H}_d(t)$. The objective is to solve for $\psi_n(t)$. Substituting Eq. (\ref{eq.III.B.03}) into Eq. (\ref{eq.III.A.05}), the equation for $\psi_n(t)$ is
\begin{equation}
\frac{d\psi_m(t)}{dt}
=
\lambda
\sum_n
\psi_n(t)
e^{(\varepsilon_n -\varepsilon_m )t}
\langle m|\mathcal{H}_d(t)|n\rangle.
\label{eq.III.B.04}
\end{equation}
We now expand $\psi_n(t)$ in powers of $\lambda$,
\begin{equation}
\psi_n(t)
=
\psi^{(0)}_n
+
\lambda
\psi^{(1)}_n(t)
+
\lambda^2
\psi^{(2)}_n(t)
+
\cdots,
\label{eq.III.B.05}
\end{equation}
where $\psi^{(r)}_n(t)$ denotes the $r$th-order approximation of $\psi_n(t)$, and $\psi^{(0)}_n$ are independent of time and determined by the initial condition $|\psi(0)\rangle$. Substituting Eq. (\ref{eq.III.B.05}) into Eq. (\ref{eq.III.B.04}) and collecting powers of $\lambda$, one obtains the recursive relation, 
\begin{equation}
\frac{d\psi_m^{(r+1)}(t)}{dt}
=
\sum_n
\psi_n^{(r)}(t)
e^{(\varepsilon_n -\varepsilon_m )t}
\langle m|\mathcal{H}_d(t)|n\rangle.
\label{eq.III.B.06}
\end{equation}
Starting from the lowest-order coefficients $\psi_n^{(0)}$, one obtains successively higher-order ones recursively using Eq. (\ref{eq.III.B.06}). Specifically, the $(r+1)$th-order coefficients are obtained by integrating the $r$th-order ones.

\subsection{Perturbative expansion of $\mathcal{T}$}
\label{sec.sub.expansion of T}

We now use the recursive relation Eq. (\ref{eq.III.B.06}) to calculate the perturbative expansion of the time evolution of the basis vectors,
\begin{equation}
|\sigma(t)\rangle
=
|\sigma^{(0)}(t)\rangle
+
\lambda
|\sigma^{(1)}(t)\rangle
+
\lambda^2
|\sigma^{(2)}(t)\rangle
+
\cdots,
\label{eq.III.C.01}
\end{equation}
where $|\sigma(t)\rangle$ is the $\sigma$-eigenvector of $\sigma^z$ at time $t$ subjected to the initial condition $|\sigma(0)\rangle=|\sigma\rangle$, and $|\sigma^{(r)}(t)\rangle$ is the $r$th-order approximation of $|\sigma(t)\rangle$. The expansion of $\mathcal{T}$ is then,
\begin{equation}
\mathcal{T} = \mathcal{T}^{(0)} + \lambda\mathcal{T}^{(1)} + \lambda^2\mathcal{T}^{(2)} + \cdots,
\label{eq.III.C.02}
\end{equation}
where 
\begin{equation}
\mathcal{T}^{(r)}
=
\sum_{\sigma=\pm1}
\langle \sigma(0) | \sigma^{(r)}(1) \rangle.
\label{eq.III.C.03}
\end{equation}

Let us denote the two values taken by the index $n$ in Eq. (\ref{eq.III.B.03}) as $+$ and $-$. The two eigenvalues of $\mathcal{H}_s$ are then denoted as $\varepsilon_+=+\varepsilon=+\sqrt{(\beta\Gamma)^2 + (2\beta J m_s)^2}$ and $\varepsilon_-=-\varepsilon$, and the corresponding eigenvectors as $|+\rangle$ and $|-\rangle$. Eq. (\ref{eq.III.B.06}) can then be written in matrix form as
\begin{equation}
\frac{d}{dt}
\left(
\begin{array}{c}
\psi_+^{(r+1)}(t) \\
\psi_-^{(r+1)}(t) \\
\end{array}
\right)
=
2\beta J
\left(
\begin{array}{cc}
\alpha m_d(t) & \gamma m_d(t) e^{-2\varepsilon t} \\
\gamma  m_d(t) e^{2\varepsilon t}& -\alpha m_d(t)\\
\end{array}
\right)
\left(
\begin{array}{c}
\psi_+^{(r)}(t) \\
\psi_-^{(r)}(t) \\
\end{array}
\right),
\label{eq.III.C.04}
\end{equation}
where we have denoted $\alpha=\langle+|\sigma^z|+\rangle=-\langle-|\sigma^z|-\rangle=\frac{2\beta J m_s}{\varepsilon}$ and $\gamma=\langle-|\sigma^z|+\rangle=\langle+|\sigma^z|-\rangle=-\frac{\beta\Gamma}{\varepsilon}$. When integrating Eq. (\ref{eq.III.C.04}), the boundary conditions are $\psi_n^{(r)}(0)=0$ for $r\ge1$.

With $\{|+\rangle,|-\rangle\}$ as basis, let us denote 
\begin{equation}
|+1(t)\rangle=
\left(
\begin{array}{c}
a_+(t) e^{\varepsilon t} \\
a_-(t) e^{-\varepsilon t}\\
\end{array}
\right)
=
\left(
\begin{array}{c}
a^{(0)}_+ e^{\varepsilon t}\\
a^{(0)}_- e^{-\varepsilon t}\\
\end{array}
\right)
+
\lambda
\left(
\begin{array}{c}
a^{(1)}_+(t) e^{\varepsilon t}\\
a^{(1)}_-(t) e^{-\varepsilon t}\\
\end{array}
\right)
+
\lambda^2
\left(
\begin{array}{c}
a^{(2)}_+(t) e^{\varepsilon t}\\
a^{(2)}_-(t) e^{-\varepsilon t}\\
\end{array}
\right)
+
\cdots
,
\label{eq.III.C.05}
\end{equation}
and a similar notation for the expansion of $|-1(t)\rangle={b_+(t)\,e^{\varepsilon t} \choose b_-(t)\,e^{-\varepsilon t}}$. This is simply rewriting Eq. (\ref{eq.III.B.03}) in vector form with $|\sigma(t)\rangle$ [c.f. Eq. (\ref{eq.III.C.01})] for $|\psi(t)\rangle$. The orthogonality conditions at $t=0$ between the two normalized eigenvectors of $\sigma^z$ gives
\begin{equation}
|a^{(0)}_{\pm}|^2+|b^{(0)}_{\pm}|^2=1,
\label{eq.III.C.06}
\end{equation}
\begin{equation}
a^{(0)}_+a^{(0)}_- + b^{(0)}_+b^{(0)}_- =0.
\label{eq.III.C.07}
\end{equation}

We now calculate $\mathcal{T}^{(1)}$. From Eq. (\ref{eq.III.C.04}), we have
\begin{equation}
\left(
\begin{array}{c}
a_+^{(1)}(t) \\
a_-^{(1)}(t) \\
\end{array}
\right)
=
2\beta J
\left(
\begin{array}{cc}
\alpha \int_0^t dt^{\prime} m_d(t^{\prime}) & \gamma \int_0^t dt^{\prime}m_d(t^{\prime}) e^{-2\varepsilon t^{\prime}}\\
\gamma \int_0^t dt^{\prime} m_d(t^{\prime}) e^{2\varepsilon t^{\prime}}& -\alpha \int_0^t dt^{\prime} m_d(t^{\prime})\\
\end{array}
\right)
\left(
\begin{array}{c}
a_+^{(0)} \\
a_-^{(0)} \\
\end{array}
\right),
\label{eq.III.C.08}
\end{equation}
for the first-order terms of $|+1(t)\rangle$. Then,
\begin{equation}
\langle +1(0) | +1^{(1)}(1)\rangle
=
\left(
\begin{array}{cc}
 a_+^{(0)}  &  a_-^{(0)} \\
\end{array}
\right)
\left(
\begin{array}{c}
a_+^{(1)}(1) e^{\varepsilon} \\
a_-^{(1)}(1) e^{-\varepsilon}\\
\end{array}
\right).
\label{eq.III.C.09}
\end{equation}
Substituting Eq. (\ref{eq.III.C.08}) into Eq. (\ref{eq.III.C.09}), and then summing with $\langle -1(0) | -1^{(1)}(1)\rangle$, one has
\begin{equation}
\mathcal{T}^{(1)}
=
2\alpha(2\beta J)\sinh\varepsilon \int_0^1 dt^{\prime} m_d(t^{\prime}),
\label{eq.III.C.10}
\end{equation}
where the orthogonality conditions Eqs. (\ref{eq.III.C.06}) and (\ref{eq.III.C.07}) have been used. Higher-orders terms are calculated similarly. In the following, we simply state the results. 

Let us introduce the notation,
\begin{equation} 
M_{s_1 \cdots s_k}
=
\int_0^1 dt_1
\,
m_d(t_{1}) e^{s_{1}2\varepsilon t_{1}}
\cdots
\int_0^{t_{k-1}} dt_k
\,
m_d(t_{k}) e^{s_{k}2\varepsilon t_{k}}.
\label{eq.III.C.11}
\end{equation} 
The subscript $s_a$ ($a\in\{1,\cdots,k\}$) indicates the sign of the exponent of $e^{s_a 2\varepsilon t_a}$ and is either $+$, or $0$, or $-$. $k$ indicates that $M_{s_1 \cdots s_k}$ is a $k$-fold integral. For example, 
\begin{equation}
M_{+0-}
=
\int_0^1 dt_1 \, m_d(t_1)e^{2\varepsilon t_1}
\int_0^{t_1} dt_2 \, m_d(t_2)
\int_0^{t_2} dt_3 \, m_d(t_3) e^{-2\varepsilon t_3}
.
\label{eq.III.C.12}
\end{equation}
With this notation, the first 5 terms in the expansion of $\mathcal{T}$ are,
\begin{align}
\mathcal{T}^{(0)} &= 2\cosh\varepsilon. \label{eq.III.C.13}\\
\mathcal{T}^{(1)} &= (2\beta J)2\alpha\sinh\varepsilon \,M_0.   \label{eq.III.C.14}\\ 
\mathcal{T}^{(2)} &=
(2\beta J)^2
\{
2\alpha^2\cosh\varepsilon \, M_{00}
+
\gamma^2
[
e^{\varepsilon}
M_{-+}
+
e^{-\varepsilon}
M_{+-}
]
\}.
\label{eq.III.C.15}\\ 
\mathcal{T}^{(3)} &=
(2\beta J)^3
\{
2\alpha^3\sinh\varepsilon\,M_{000} 
\nonumber\\ 
&
+ \alpha\gamma^2 [ e^{\varepsilon}(M_{0-+}-M_{-0+} + M_{-+0})
 -e^{-\varepsilon} ( M_{0+-}-M_{+0-}  + M_{+-0})
]
\}.
\label{eq.III.C.16}\\ 
\mathcal{T}^{(4)} &=
(2\beta J)^4
\{
2\alpha^4\cosh\varepsilon\,M_{0000} 
+ 
\gamma^4[e^{\varepsilon}M_{-+-+} + e^{-\varepsilon}M_{+-+-}]
\nonumber\\ 
&
+
\alpha^2\gamma^2
[
e^{\varepsilon}(M_{-+00}-M_{-0+0}+M_{0-+0}-M_{0-0+}+M_{00-+}+M_{-00+})
\nonumber\\ 
&
+e^{-\varepsilon}(M_{+-00}-M_{+0-0}+M_{0+-0}-M_{0+0-}+M_{00+-}+M_{+00-})
]\}.
\label{eq.III.C.17}
\end{align}

\section{Ground-state energy: Non-extensive terms}
\label{sec.E0}

The formalism developed in Sec. \ref{sec.nonstaticity} will now be applied to calculate the ground-state energy $E_0$ beyond the extensive term given by the static approximation. In Sec. \ref{sec.sub.E0 2nd order}, we first show that the $N^0$ term obtained by the Holstein-Primakoff transform is reproduced by expanding $\mathcal{T}$ to second-order. In Sec. \ref{sec.sub.E0 4th order}, the subsequent $N^{-1}$ term is calculated by further expanding $\mathcal{T}$ to the fourth-order.

\subsection{Second-order approximation and $N^0$ term of $E_0$}
\label{sec.sub.E0 2nd order}

We first consider second-order approximation. Eq. (\ref{eq.III.C.02}) becomes, 
\begin{equation}
\mathcal{T}
\stackrel{\mathrm{2nd}}{\longrightarrow}
\mathcal{T}^{(0)}
+
\lambda
\mathcal{T}^{(1)}
+
\lambda^2
\mathcal{T}^{(2)},
\label{eq.IV.A.01}
\end{equation}
where $\stackrel{\mathrm{2nd}}{\longrightarrow}$ denotes second-order approximation. Substituting Eqs. (\ref{eq.III.C.13})-(\ref{eq.III.C.15}) into Eq. (\ref{eq.IV.A.01}), inserting the latter into Eq. (\ref{eq.II.B.04}), and collecting together the same powers of $\lambda$, the partition function becomes, 
\begin{align}
Z
\stackrel{\mathrm{2nd}}{\longrightarrow}
&
\,
e^{-\beta N f_s} \int Dm_d(t)    
\nonumber\\
&
\times
\exp
\left[
-
\lambda N
\,
(2\beta J)
\,
m_s
\left(
1-
\frac{2\beta J}{\varepsilon}\tanh\varepsilon
\right)
M_0
\right]
\nonumber\\
&
\times
\exp 
\left[
\lambda^2 N
\left(
-\beta J
\int_0^1 dt \, m_d^2(t)
+
\frac{(2\beta J)^2 \gamma^2}{2\cosh\varepsilon}[e^{\varepsilon}M_{-+}+e^{-\varepsilon}M_{+-}]
\right)
\right]
\nonumber\\
&
\times
\exp 
\left[
\lambda^2 N
\,
(2\beta J)^2
\alpha^2
\,
\left(
M_{00}
-
\frac{1}{2}
\tanh^2\varepsilon
(M_0)^2
\right)
\right]
\times
e^{O(N\lambda^3)},
\label{eq.IV.A.02}
\end{align}
where $\int Dm_d(t)$ denotes summing over all $m_d(t)$, the non-static paths \cite{note.notation for measure}. The order $\lambda$ term in the second line vanishes because of the stationary condition Eq. (\ref{eq.II.C.05}). The exponent in the fourth line is in fact $O(\mathrm{sech}^2\varepsilon)$ and can be dropped in the limit $\beta\rightarrow\infty$ \cite{note.concerning sech2e term}. It remains to integrate over the third line of Eq. (\ref{eq.IV.A.02}). From $\lambda^2 N$, we also see that $\lambda$ needs to be $1/\sqrt{N}$ in order for the integral not to diverge or to vanish as $N\rightarrow\infty$. 

The path integral is performed by expanding $m_d(t)$ in Fourier series, 
\begin{equation}
m_d(t)
=
\sum_{n=-\infty}^{\infty}
c_n
\,
e^{i 2\pi n t}.
\label{eq.IV.A.03}
\end{equation}
As $m_d(t)$ is real, $c_n^*=c_{-n}$. One has $\int_0^1 dt\, m^2_d(t)=\sum_{n=-\infty}^{\infty}c_n c_{-n}$, and,   
\begin{equation}
e^{\varepsilon} M_{-+} + e^{-\varepsilon} M_{+-} 
=
4\varepsilon\sinh\varepsilon
\sum_{n=-\infty}^{\infty}
\frac{c_nc_{-n}}{(2\pi n)^2+(2\varepsilon)^2}.
\label{eq.IV.A.04}
\end{equation}
Eq. (\ref{eq.IV.A.02}) becomes,
\begin{equation}
Z
\stackrel{\mathrm{2nd}}{\longrightarrow}
C
\,
e^{-\beta N f_s}
\int
dc_0
\prod_{n=1}^{\infty}
dc_n dc_n^*
\,
\exp
\left(
-\beta J \sum_{n=-\infty}^{\infty}g_n c_n c_{-n}
\right),
\label{eq.IV.A.05}
\end{equation}
where $g_n=1-\frac{g}{(2\pi n)^2+(2\varepsilon)^2}$ with $g=\frac{8\Gamma^2 J \beta^3\tanh\varepsilon}{\varepsilon}$, and $dc_ndc_n^*$ means $d\mathrm{Re}(c_n)d\mathrm{Im}(c_n)$\cite{note.complex gaussian integral formula}. The constant $C=\sqrt{\frac{\beta J}{\pi}}\prod_{n=1}^{\infty}(\frac{2\beta J}{\pi})$. Performing the gaussian integrals, and using the formula
\begin{equation}
\prod_{n=1}^{\infty}
\left(
1+
\frac{z_1}{n^2+z_2}
\right)
=
\frac{\sqrt{z_2}}{\sqrt{z_1+z_2}}
\frac{\sinh\pi\sqrt{z_1+z_2}}{\sinh\pi\sqrt{z_2}},
\label{eq.IV.A.06}
\end{equation}
to evaluate the resulting infinite product sequence, we obtain
\begin{equation}
Z
\stackrel{\mathrm{2nd}}{\longrightarrow}
e^{-\beta N f_s}
\frac{\sinh \varepsilon}{\sinh \sqrt{\varepsilon^2 -\frac{g}{4} }}.
\label{eq.IV.A.07}
\end{equation}
Inserting Eq. (\ref{eq.IV.A.07}) into Eq. (\ref{eq.II.B.01}), the ground-state energy given by second-order approximation is
\begin{equation}
E_0
\stackrel{\mathrm{2nd}}{\longrightarrow}
Nf_s
+
\sqrt{\Gamma^2+(2Jm_s)^2 - \frac{2\Gamma^2 J}{\sqrt{\Gamma^2+(2Jm_s)^2}}}
-
\sqrt{\Gamma^2+(2Jm_s)^2}.
\label{eq.IV.A.08}
\end{equation}
Substituting the solution for $m_s$ Eq. (\ref{eq.II.C.06}), we have
\begin{equation}
E_0
\stackrel{\mathrm{2nd}}{\longrightarrow}
\left\{
\begin{array}{ccc}
-N \Gamma + \sqrt{\Gamma(\Gamma-2J)}-\Gamma  & \mathrm{for} & \Gamma \ge2J, \\
-N \frac{(2J)^2+\Gamma^2}{4J}+\sqrt{(2J)^2-\Gamma^2} -2J  & \mathrm{for} & \Gamma <2J. \\
\end{array}
\right.
\label{eq.IV.A.09}
\end{equation}
Comparing with Eq. (\ref{eq.II.A.03}), we see that we have recovered the order $N^0$ term obtained by Holstein-Primakoff transform. 

\subsection{Fourth-order approximation and $N^{-1}$ term of $E_0$}
\label{sec.sub.E0 4th order}

We now consider fourth-order approximation. Eq. (\ref{eq.III.C.02}) now becomes 
\begin{equation}
\mathcal{T}
\stackrel{\mathrm{4th}}{\longrightarrow}
\mathcal{T}^{(0)}
+
\lambda
\mathcal{T}^{(1)}
+
\lambda^2
\mathcal{T}^{(2)}
+
\lambda^3
\mathcal{T}^{(3)}
+
\lambda^4
\mathcal{T}^{(4)}
,
\label{eq.IV.B.01}
\end{equation}
where $\stackrel{\mathrm{4th}}{\longrightarrow}$ denotes fourth-order approximation. Following similar derivation steps that have led from Eq. (\ref{eq.IV.A.01}) to Eq. (\ref{eq.IV.A.05}), the partition function is now
\begin{equation}
Z
\stackrel{\mathrm{4th}}{\longrightarrow}
C
\,
e^{-\beta N f_s}
\int
dc_0
\prod_{n=1}^{\infty}
dc_n dc_n^*
\,
\left[
1
+
\frac{1}{N}
\left(
V_4
+
\frac{1}{2}
\left(
V_3
\right)^2
\right)
\right]
\,
\exp
\left(
-\beta J \sum_{n=-\infty}^{\infty}g_n c_n c_{-n}
\right)
,
\label{eq.IV.B.02}
\end{equation}
where
\begin{align}
V_3
=
&
L_3
-
L_1L_2
+
\frac{1}{3}\left(L_1\right)^3
,
\label{eq.IV.B.03} \\
V_4
=
&
L_4
-
L_1L_3
-
\frac{1}{2}
\left(L_2\right)^2
+
\left(L_1\right)^2L_2
-
\frac{1}{4}\left(L_1\right)^4
,
\label{eq.IV.B.04}
\end{align}
and $L_i$ denotes $\frac{\mathcal{T}^{(i)}}{\mathcal{T}^{(0)}}$\cite{note.reason for dropping V3/sqrtN term}. The Fourier expansions of $V_3$ and $V_4$ are obtained by first substituting Eq. (\ref{eq.IV.A.03}) into $\mathcal{T}^{(i)}$ given by Eqs. (\ref{eq.III.C.14}) to (\ref{eq.III.C.17}) and then inserting the results into Eqs. (\ref{eq.IV.B.03}) and (\ref{eq.IV.B.04}). The expansions of $\mathcal{T}^{(1)}$ and $\mathcal{T}^{(2)}$ have already been calculated in Sec. \ref{sec.sub.E0 2nd order}\cite{note.instructions for L1 and L2}. The expansions of $\mathcal{T}^{(3)}$ and $\mathcal{T}^{(4)}$ are given in Appendix \ref{sec.appendix.fourier expansions of T3 and T4}, and that of certain terms arising from $\left(\mathcal{T}^{(3)}\right)^2$ in Appendix \ref{sec.appendix.calculation of half V3 square}.

We next integrate over $V_4$ and $\frac{1}{2}(V_3)^2$ with the gaussian function, substitute the resulting $Z$ into Eq. (\ref{eq.II.B.01}), and then keep only those terms that do not vanish in the limit $\beta\rightarrow\infty$. Along the way, the formula
\begin{equation}
\sum_{n=1}^{\infty}
\frac{1}{z_1 n^2 + z_2}
=
-\frac{1}{2z_2}
+
\frac{\pi}{2\sqrt{z_1z_2}}
\,
\mathrm{coth}
\left(
\pi\sqrt{\frac{z_2}{z_1}}
\right)
,
\label{eq.IV.B.05}
\end{equation}
is used to evaluate some of the summations that appear and check their powers of $\beta$. One then gets
\begin{equation}
E_0
\stackrel{\mathrm{4th}}{\longrightarrow}
Nf_s
+
\frac{\sqrt{\varepsilon^2 -\frac{g}{4} }-\varepsilon}{\beta}
-
\frac{1}{N}
\left(
\frac{z_4^1 + z_4^2 + z_3^1 +z_3^2}{\beta}
\right)
.
\label{eq.IV.B.06}
\end{equation}
The first two terms of Eq. (\ref{eq.IV.B.06}) have already been obtained in Eq. (\ref{eq.IV.A.08}). For the $N^{-1}$ term, we have \cite{note.explaining origin of each of the terms making up N-1 term} 
\begin{align}
z_4^1
&
=
(2\beta J)^2 8\varepsilon \gamma^4 \tanh\varepsilon
\left[
\sum_{n=1}^{\infty}
\frac{1}{g_n[n]}
\right]
\left[
\sum_{n=1}^{\infty}
\frac{1}{g_n[n]}
-
4(2\varepsilon)^2
\sum_{n=1}^{\infty}
\frac{1}{g_n[n]^2}
\right],
\label{eq.IV.B.07}\\[15pt]
z_4^2
&
=
(2\beta J)^2 16\varepsilon\alpha^2\gamma^2\tanh\varepsilon
\left[
\left(
\sum_{n=1}^{\infty}
\frac{1}{g_n [n]}
\right)^2
+
(2\varepsilon)^2
\sum_{n=1}^{\infty}
\sum_{m=1}^{\infty}
\left(
\frac{1}{g_n [n]}
+
\frac{1}{g_m [m]}
\right)
\frac{1}{[n+m][n-m]}
\right.
\nonumber\\
&
+
\left.
(2\varepsilon)^2
\left(
2g
+
3(2\varepsilon)^2
\right)
\sum_{n=1}^{\infty}
\sum_{m=1}^{\infty}
\frac{1}{g_n g_m[n][m][n+m][n-m]}
\right],
\label{eq.IV.B.08}\\[15pt]
z_3^1
&
=
\frac{(2\beta J)^4 4\alpha^2 \gamma^4 \tanh^2\varepsilon}{\beta J g_0}
\left[
\sum_{n=1}^{\infty}
\frac{1}{g_n[n]}
+
8\varepsilon^2
\sum_{n=1}^{\infty}
\frac{1}{g_n[n]^2}
\right]^2
,
\label{eq.IV.B.09}\\[15pt]
z_3^2
&
=
(2\beta J)^3 8(2\varepsilon)^4\alpha^2\gamma^4\tanh^2\varepsilon
\sum_{n=1}^{\infty}\sum_{m=1}^{\infty}
\frac{1}{g_{n+m}g_ng_m [n+m]^2 }
\left[
\frac{1}{[n]}
+
\frac{1}{[m]}
+
\frac{(2\pi n)(2\pi m)+(2\varepsilon)^2}{[n][m]}
\right]^2,
\label{eq.IV.B.10}
\end{align}
where $[n]=(2\pi n)^2+(2\varepsilon)^2$. For Eqs. (\ref{eq.IV.B.07}) and (\ref{eq.IV.B.09}), using partial fractions to simplify the summands and then using Eq. (\ref{eq.IV.B.05}), one obtains
\begin{align}
z_4^1
&
=
\frac{(2\beta J)^2(2\varepsilon)^2 \gamma^4}{g}
\left[
\frac{1}{\sqrt{(2\varepsilon)^2-g}}
+
\frac{1}{(2\varepsilon)^2-g}
\left(
\frac{g}{8\varepsilon}-2\varepsilon
\right)
\right],
\label{eq.IV.B.11}\\[10pt]
z_3^1
&
=
\frac{m_s^2(2\beta J)^6\gamma^4}{\beta J g_0}
\left[
\frac{4}{g^2}
+
\frac{1}{(2\varepsilon)^2-g}
\left[
\frac{1}{(2\varepsilon)^2}
+
\frac{4}{g}
+
\left(\frac{4\varepsilon}{g}\right)^2
\right]
-
\frac{2}{g\sqrt{(2\varepsilon)^2-g}}
\left[
\frac{1}{\varepsilon}
+
\frac{8\varepsilon}{g}
\right]
\right]
,
\label{eq.IV.B.12}
\end{align}
where terms of order $\beta^0$ and smaller have been dropped.

\begin{figure}[h]
\begin{center}
\includegraphics[scale=1.0]{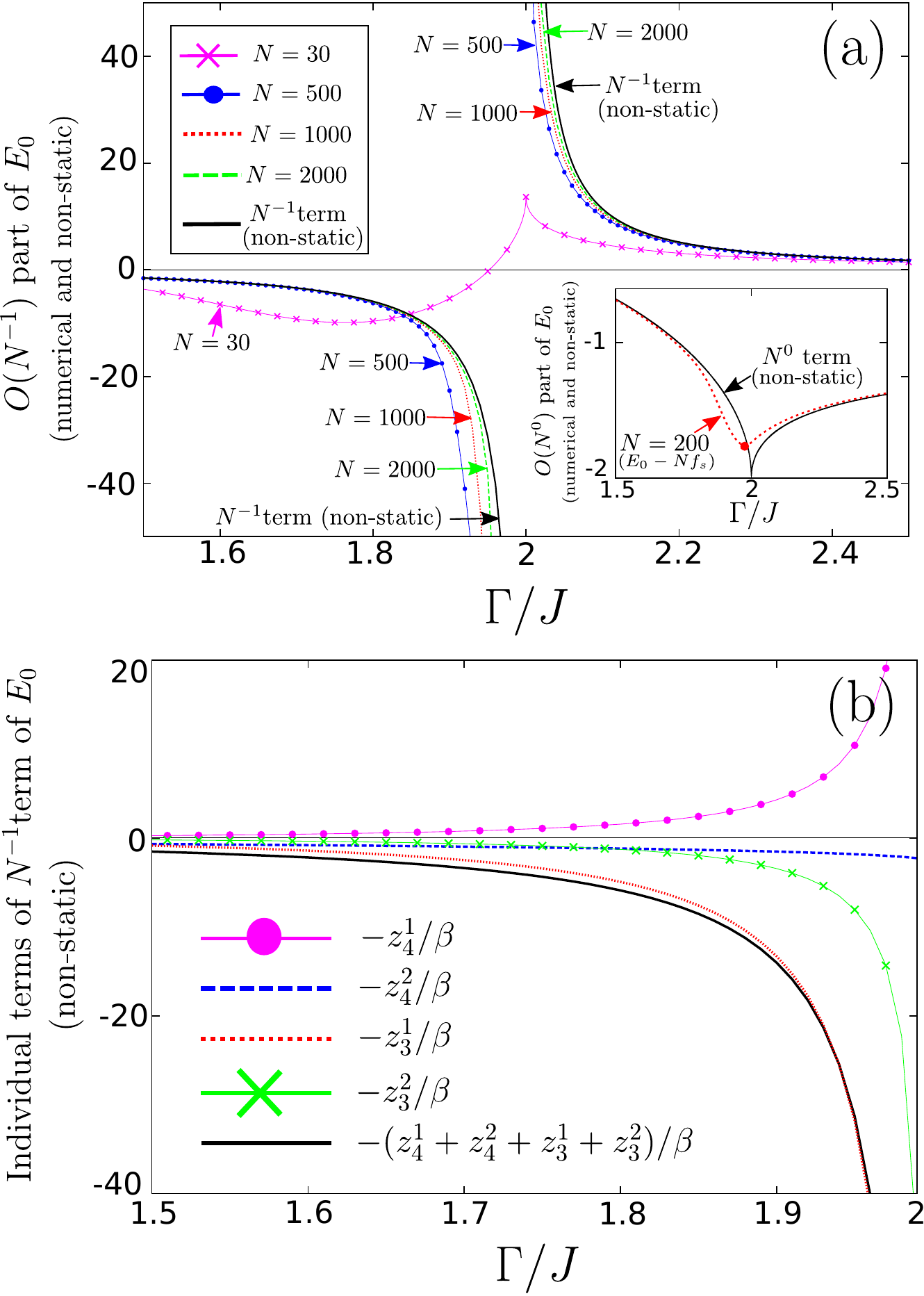}
\caption{(a) Comparing the $N^{-1}$ term of $E_0$ obtained using non-static fourth-order approximation (black solid line) with numerical results ($N=30,500,1000$, and $2000$). Details of the latter are given in the text. Inset: To understand the difference in signs within the two phases, the $N^0$ term in Eq. (\ref{eq.IV.A.09}) is compared to $(E_0-Nf_s)$ for $N=200$. The latter lies below (above) the former in the ferromagnetic (paramagnetic) phase. (b) Contributions by the individual terms appearing in Eq. (\ref{eq.IV.B.06}) to the total $N^{-1}$ term, in the ferromagnetic phase.}
\label{fig.inverse N term of E0}
\end{center}
\end{figure}

Let us first consider the paramagnetic phase. In this phase, only $z_4^1$ contributes because the prefactors of $z_4^2$, $z_3^1$, and $z_3^2$ contain $m_s$ in $\alpha$ and hence vanishes. From Eq. (\ref{eq.IV.B.11}), one gets
\begin{equation}
z_4^1
=
\beta
\left[
\frac{\Gamma J}{\sqrt{\Gamma(\Gamma-2J)}}
-
\frac{J(2\Gamma-J)}{2(\Gamma-2J)}
\right]
\,\,\,\,
\mathrm{for}
\,\,
\Gamma \ge2J
.
\label{eq.IV.B.13}
\end{equation}
In Fig. \ref{fig.inverse N term of E0}(a), we plotted $-\frac{z_4^1}{\beta}$ [labelled `$N^{-1}$ term (non-static)'] in the region $\Gamma\ge 2J$. To check the correctness of our result, we compared it to numerical calculations. The Hamiltonian Eq. (\ref{eq.II.A.01}) is diagonalized in the sector with total angular momentum $N/2$ to obtain $E_0$. The $N^1$ and $N^0$ terms [given by Eq. (\ref{eq.IV.A.09})] are subtracted away from $E_0$ and the result is rescaled by multiplying by $N$. The curves for $N=30$, 500, 1000, and 2000 calculated in this way are compared to $-\frac{z_4^1}{\beta}$ in Fig. \ref{fig.inverse N term of E0}(a). 

In the ferromagnetic phase, all four terms $z_4^1$, $z_4^2$, $z_3^1$, and $z_3^2$ contribute. From Eqs. (\ref{eq.IV.B.11}) and (\ref{eq.IV.B.12}), $z_4^1$ and $z_3^1$ become
\begin{align}
z_4^1
=
&
\,
\beta
\left[
\frac{\Gamma^2}{2\sqrt{(2J)^2-\Gamma^2}}
-
\frac{\Gamma^2[(4J)^2-\Gamma^2]}{16J[(2J)^2-\Gamma^2]}
\right]
\,\,\,\,
\mathrm{for}
\,\,
\Gamma<2J
,
\label{eq.IV.B.14}\\
z_3^1
=
&
\,
\beta 
\left[ 
J
+ 
\frac{J\Gamma^4 }{(2J)^2-\Gamma^2}\left[\frac{1}{16J^2} + \frac{1}{\Gamma^2} + \left(\frac{2J}{\Gamma^2}\right)^2 \right]
-
\frac{J\Gamma^2}{\sqrt{(2J)^2-\Gamma^2}}\left[\frac{1}{2J} + \frac{4J}{\Gamma^2}\right]  
\right]
\,\,\,\,
\mathrm{for}
\,\,
\Gamma<2J
.
\label{eq.IV.B.15}
\end{align}
Due to the double summations appearing in $z_4^2$ and $z_3^2$, these two terms have be evaluated numerically\cite{note.concerning computing double sums of z42 and z32 numerically}. Fig. \ref{fig.inverse N term of E0}(a) shows the curve of $-\frac{1}{\beta}(z_4^1+z_4^2+z_3^1+z_3^2)$ in the region $\Gamma<2J$. Results from numerical diagonalization of the Hamiltonian are again shown for comparison. 

From Fig. \ref{fig.inverse N term of E0}(a) we see that the $N^{-1}$ term of $E_0$ is negative in the ferromagnetic phase and positive in the paramagnetic one. To elucidate on this point, the inset of Fig. \ref{fig.inverse N term of E0}(a) compares the $N^0$ term [c.f. Eq. (\ref{eq.IV.A.09})] to $E_0-Nf_s$ for $N=200$ where $E_0$ is computed by numerical diagonalization and $Nf_s$ is the $N^1$ term. The signs of the $N^{-1}$ term in the two phases can be evinced by noting that the curve of $N=200$ lies below that of the $N^{0}$ term in the ferromagnetic phase and above it in the paramagnetic one.

Fig. \ref{fig.inverse N term of E0}(a) also shows that the $N^{-1}$ term diverges at the critical point. This divergence can be understood by examining the rate at which the minimum point of $E_0-Nf_s$ (indicated by a red solid circle on the $N=200$ curve of the inset) converges towards the critical value of -2 at $\Gamma=2J$ as $N$ increases. We found numerically that the difference between the critical and finite-$N$ value decreases as $N^{-0.33}$. Upon rescaling by multiplying by $N$, this decrease is turned into an increase that scales as $N^{0.67}$, thereby accounting for the divergence.

Fig. \ref{fig.inverse N term of E0}(b) shows the individual terms $z_4^1$, $z_4^2$, $z_3^1$, and $z_3^2$ that make up the $N^{-1}$ term, in the ferromagnetic phase. The terms $z_4^1$ and $z_3^2$ approximately cancel each other, while the magnitude of $z_4^2$ is the smallest among the four. Although the greatest contribution might appear to come from just $z_3^1$, it is important to rigorously sum up all four terms to arrive at the proper result. Away from the critical point, one sees that both $z_4^2$ and $z_3^1$ actually contribute equally much to the final curve. In the vicinity of the critical point, $z_3^1$ alone will diverge too quickly if the contribution of the positive and weakly-diverging $z_4^1+z_3^2$ is not accounted for. 

We conclude this section on ground-state energy with a comment on $c_0$, the zero mode of the Fourier expansion Eq. (\ref{eq.IV.A.03}). Throughout our calculations, we have included $c_0$ in the expansion although it is not \emph{a priori} evident whether this is necessary since one might imagine that it can also be absorbed into the static term $m_s$. Indeed, one still obtains the same result for the $N^0$ term in Sec. \ref{sec.sub.E0 2nd order} if $c_0$ is excluded because this simply introduces a $\beta$-independent multiplicative factor to $Z$ that ultimately vanishes when taking the limit $\beta\rightarrow\infty$ in Eq. (\ref{eq.II.B.01}). The result for the $N^{-1}$ term in the paramagnetic phase is also not affected because $c_0$ is not involved in the derivation of $z_4^1$. However, our analysis of the $N^{-1}$ term in the ferromagnetic phase shows that $c_0$ does play a role. The term $z_3^1$ originates from the coupling of $c_0$ to other non-zero modes when taking the square in $\frac{1}{2}(V_3)^2$. (This can be discerned from the presence of $g_0$ in the denominator of the prefactor of Eq. (\ref{eq.IV.B.09}).) Excluding $z_3^1$ from the $N^{-1}$ term will result in disagreement between non-static results and finite-$N$ numerical calculations. This insight into the importance of $c_0$ based on an ordered system will be useful when applying the non-static framework to disordered ones. For the disordered models, we no longer have means to check our non-static results since accurate ground-state energies of large-sized systems are difficult to obtain numerically.

\section{Energy gap using parity operator}
\label{sec.gap 1}

\subsection{Formulation of first excited-state energy $E_1$}
\label{sec.sub.gap 1 formulate}

In this section, we derive the first of two formulae for calculating the first excited-state energy $E_1$. The form of these formulae are similar to that of Eq. (\ref{eq.II.B.01}) for $E_0$, and so are amenable to path integral calculations. 

Consider an operator $Q$ with the property, 
\begin{equation}
Q^2=I,
\label{eq.V.A.01}
\end{equation}
where $I$ is the identity operator. $Q$ has eigenvalues $+1$ and $-1$. Let us call $Q$ the parity operator in analogy with the reflection operation in one-dimensional space. An eigenvector of $Q$ with eigenvalue +1 ($-1$) is said to have even (odd) parity.  

Let $|E_n^a\rangle$ denote an eigenvector of the Hamiltonian $H$ with energy $E_n$, where $a$ denotes the rest of the quantum numbers required to specify the state. Suppose $Q$ commutes with $H$, i.e., $[H,Q]=0$. Then $H$ and $Q$ can be simultaneously diagonalized such that the energy eigenvectors $|E_n^a\rangle$ have either even or odd parity. Suppose further that $Q|E_0^a\rangle=|E_0^a\rangle$ for any $a$, and that $Q|E_1^a\rangle=-|E_1^a\rangle$ for any $a$. Then,
\begin{equation}
e^{-\beta H} =  e^{-\beta E_0}\left( \sum_{a} |E_0^a\rangle\langle E_0^a| \right)+  e^{-\beta E_1} \left(\sum_{a} |E_1^a\rangle\langle E_1^a|\right) + \cdots,
\label{eq.V.A.02}
\end{equation}
and\begin{equation}
Qe^{-\beta H} =  e^{-\beta E_0}\left( \sum_{a} |E_0^a\rangle\langle E_0^a| \right)-  e^{-\beta E_1} \left(\sum_{a} |E_1^a\rangle\langle E_1^a| \right)+ \cdots,
\label{eq.V.A.03}
\end{equation}
where $\sum_a$ denotes summing over the basis of the, possibly degenerate, energy level. Subtracting Eq. (\ref{eq.V.A.03}) from Eq. (\ref{eq.V.A.02}) and then taking the trace, we have,
\begin{equation}
\mathrm{Tr}
\left(
e^{-\beta H}-Qe^{-\beta H} 
\right)
=  
2 d_1 e^{-\beta E_1} 
\left[
1 + O\left(e^{-\beta (E_2-E_1)}\right)
\right],
\label{eq.V.A.04}
\end{equation}
where $d_1$ is the degeneracy of the first excited energy level. Define 
\begin{equation}
Z_Q=\mathrm{Tr}\left(Qe^{-\beta H}\right).
\label{eq.V.A.05}
\end{equation}
Then, 
\begin{equation}
E_1=\lim_{\beta\rightarrow \infty}-\frac{1}{\beta}\ln \left(Z-Z_{Q}\right).
\label{eq.V.A.06}
\end{equation}
Note that the subtraction $Z-Z_{Q}$ must be performed before taking the limit $\beta\rightarrow\infty$; otherwise, the result is zero.

In our derivation, we have used the following conditions: 
\begin{enumerate}
\item Existence of a parity operator $Q$ that commutes with $H$.

\item All the states in the ground-state energy level have even parity, and all the states in the first excited-state energy level have odd parity.
\end{enumerate}

\subsection{Choice of $Q$ and the single spin trace $\mathcal{T}_Q$}
\label{sec.sub.TQ}

The relation Eq. (\ref{eq.V.A.06}) is general and not specific to any particular model. Let us restrict ourselves now to the ferromagnetic model Eq. (\ref{eq.II.A.01}). The operator
\begin{equation}
Q=\prod_{i=1}^{N} \sigma_i^x
\label{eq.V.B.01}
\end{equation}
satisfies Eq. (\ref{eq.V.A.01}) and commutes with $H$, and can serve as the parity operator. However, Eq. (\ref{eq.V.A.06}) is not valid for all $\Gamma$ and $J$. For instance, in the limit $N\rightarrow\infty$, the ground-state in the ferromagnetic phase is doubly-degenerate and spanned by a basis vector with even parity and another with odd parity; condition 2 is therefore not satisfied. In Appendix \ref{sec.appendix.ZQ in para phase}, we show that condition 2 is satisfied in the paramagnetic phase.

Inserting Eq. (\ref{eq.V.B.01}) into Eq. (\ref{eq.V.A.05}) and following the same steps as in Sec. \ref{sec.sub.path integral representation}, the path integral representation of $Z_Q$ is  
\begin{equation}
\left(Z_Q\right)_M=
\left(\sqrt{\frac{\beta J N}{\pi M}}\right)^{M}
\prod_{\kappa=0}^{M-1}
\int_{-\infty}^{\infty} dm_{\kappa}
\exp\left(-\frac{\beta J N}{M}\sum_{\kappa=0}^{M-1}m_{\kappa}^2\right)
\left(
\sum_{\sigma=\pm1}
\langle\sigma
|\sigma^x
\left[
\prod_{\kappa=0}^{M-1}
e^{\frac{1}{M}(\beta \Gamma \sigma^x + 2\beta J m_{\kappa} \sigma^z)}
\right]
|
\sigma\rangle
\right)^N
.
\label{eq.V.B.02}
\end{equation}
$Z_M$ and $(Z_Q)_M$ differ in the single spin trace,
\begin{align}
\mathcal{T}_Q
&
=
\sum_{\sigma=\pm1}
\langle \sigma |
\sigma^x
\left[
\prod_{\kappa=0}^{M-1}
e^{\frac{1}{M} ( \beta \Gamma \sigma^x + 2\beta Jm_{\kappa} \sigma^z  )  }
\right]
| \sigma \rangle
\label{eq.V.B.03} \\
&
=
\sum_{\sigma=\pm1}
\langle \sigma(0)|
\sigma^x
| \sigma(1) \rangle,
\label{eq.V.B.04}
\end{align}
where one multiplies the spinor $|\sigma(1)\rangle$ by the matrix $\sigma^x$ before taking the inner product with $\langle \sigma(0)|\,$\cite{note.form of sigmax in energy basis}. Expanding $\mathcal{T}_Q$ perturbatively,
\begin{equation}
\mathcal{T}_Q
=
\mathcal{T}_Q^{(0)}
+
\lambda
\mathcal{T}_Q^{(1)}
+
\lambda^2
\mathcal{T}_Q^{(2)}
+
\cdots,
\label{eq.V.B.05}
\end{equation}
the $r$th-order term
\begin{equation}
\mathcal{T}_Q^{(r)}
=
\sum_{\sigma=\pm1}
\langle \sigma(0)
|\sigma^x|
\sigma^{(r)}(1)\rangle
\label{eq.V.B.06}
\end{equation}
is calculated by repeating the steps of Sec. \ref{sec.sub.expansion of T}. We simply state the results for the first 3 terms of the expansion,
\begin{align}
\mathcal{T}_Q^{(0)}
&
=
-2\gamma \sinh\varepsilon.
\label{eq.V.B.07}\\
\mathcal{T}_Q^{(1)}
&
=
(2\beta J)
\,
\alpha\gamma
\left[
-2\cosh\varepsilon \,M_0
+
e^{\varepsilon}M_-
+
e^{-\varepsilon}
M_+
\right].
\label{eq.V.B.08}\\
\mathcal{T}_Q^{(2)}
&
=
(2\beta J)^2
\{
-2\alpha^2\gamma \sinh\varepsilon \, M_{00} - \gamma^3 (e^{\varepsilon}M_{-+}-e^{-\varepsilon}M_{+-})
\nonumber\\
&
+
\alpha^2\gamma
[
e^{\varepsilon}(M_{0-} - M_{-0})
+
e^{-\varepsilon}(M_{+0} - M_{0+})
]
\}.
\label{eq.V.B.09}
\end{align}

\subsection{Calculation of $Z_Q$ in the paramagnetic phase}
\label{sec.sub.ZQ}

\subsubsection{Static approximation}
\label{sec.sub.sub.ZQ static}

We first consider static approximation Eq. (\ref{eq.II.C.01}). The single spin trace becomes
\begin{equation}
\mathcal{T}_Q
\stackrel{\mathrm{s.a.}}{\longrightarrow}
\mathcal{T}_Q^{(0)}.
\label{eq.V.C.01}
\end{equation}
Inserting Eqs. (\ref{eq.II.C.01}) and (\ref{eq.V.B.07}) into Eq. (\ref{eq.V.B.02}), one has
\begin{equation}
Z_Q
\stackrel{\mathrm{s.a.}}{\longrightarrow}
\mathrm{const.}
\times
\int
dm_s
\exp
(-\beta N \hat{f}_s),
\label{eq.V.C.02}
\end{equation}
where 
\begin{equation}
\hat{f}_s
=
J m_s^2
-
\frac{1}{\beta}
\ln
\left(
\frac{2\beta\Gamma \sinh\varepsilon}{\varepsilon}
\right). 
\label{eq.V.C.03}
\end{equation}
In the limit $N\rightarrow\infty$, Eq. (\ref{eq.V.C.02}) is evaluated using the method of steepest descent. The stationary condition $\partial \hat{f}_s/\partial m_s=0$ gives
\begin{equation}
m_s
\left(
1
-
\frac{2J\left(\mathrm{coth}\,\varepsilon -\frac{1}{\varepsilon}\right)}{\sqrt{\Gamma^2+(2Jm_s)^2}}
\right)
=
0. 
\label{eq.V.C.04}
\end{equation}
In the limit $\beta\rightarrow\infty$, the solution is once again given by Eq. (\ref{eq.II.C.06}). As the formula Eq. (\ref{eq.V.A.06}) is applicable to the ferromagnetic model only in the paramagnetic phase, we shall be concerned only with the solution $m_s=0$. 

\subsubsection{Second-order approximation}
\label{sec.sub.sub.ZQ 2nd}

We now make the non-static ansatz Eq. (\ref{eq.III.B.01}). At second-order approximation, the single spin trace becomes
\begin{equation}
\mathcal{T}_Q
\stackrel{\mathrm{2nd}}{\longrightarrow}
\mathcal{T}_Q^{(0)}
+
\lambda
\mathcal{T}_Q^{(1)}
+
\lambda^2
\mathcal{T}_Q^{(2)}
.
\label{eq.V.C.05}
\end{equation}
Inserting Eqs. (\ref{eq.III.B.01}) and (\ref{eq.V.C.05}) into Eq. (\ref{eq.V.B.02}), one has
\begin{equation}
Z_Q
\stackrel{\mathrm{2nd}}{\longrightarrow}
e^{-\beta N \hat{f}_s}
\int Dm_d(t)
\exp
\left(
-\beta J
\int_0^1 dt \, m^2_d(t)
+
\frac{(2\beta J)^2\gamma^2}{2\sinh \varepsilon}
[
e^{\varepsilon} M_{-+}
-
e^{-\varepsilon} M_{+-}
]
\right)
\times
e^{O(N\lambda^3)}
,
\label{eq.V.C.06}
\end{equation}
where we have used $m_s=0$ and kept only those terms that do not vanish in the paramagnetic phase, and also $\lambda=1/\sqrt{N}$. 

The path integral is performed by once again expanding $m_d(t)$ in Fourier series. Here, the boundary condition of $Z_Q$ is different from that of $Z$. In the conventional $Z=\sum_{\sigma} \langle \sigma|e^{-\beta H}|\sigma\rangle$, the boundary condition is periodic because one starts at $|\sigma\rangle$ and ends at the same state $\langle \sigma|$. In $Z_Q$, however, the operator $Q$ flips the end state $\langle\sigma|$. The paths $m_d(t)$ therefore needs to start and end at opposite points, i.e.,
\begin{equation}
m_d(0)=-m_d(1).
\label{eq.V.C.07}
\end{equation}
The Fourier expansion respecting this boundary condition is then
\begin{equation}
m_d(t)
=
\sideset{}{'}
\sum_{n}
c_n
\,
e^{i\pi n t}
,
\label{eq.V.C.08}
\end{equation}
where the dummy index $n$ in $\sum_n^{'}$ runs over all positive and negative odd integers. One has $\int_0^1 dt \,m^2_d(t)=\sum_n^{'} c_n c_{-n}$ and 
\begin{equation}
e^{\varepsilon}
M_{-+}
-
e^{-\varepsilon}
M_{+-}
=
4\varepsilon\cosh \varepsilon
\sideset{}{'}
\sum_n
\frac{c_n c_{-n}}{(\pi n)^2+(2\varepsilon)^2}
.
\label{eq.V.C.09}
\end{equation}
Eq. (\ref{eq.V.C.06}) becomes
\begin{equation}
Z_Q
\stackrel{\mathrm{2nd}}{\longrightarrow}
\hat{C}
\,
e^{-\beta N \hat{f}_s}
\int
\sideset{}{'}
\prod_{n}
dc_n dc_n^*
\exp
\left(
-\beta J
\sideset{}{'}
\sum_n
\hat{g}_n
c_n c_{-n}
\right),
\label{eq.V.C.10}
\end{equation}
where $\hat{g}_n=1-\frac{\hat{g}}{(\pi n)^2+(2\varepsilon)^2}$ with $\hat{g}=\frac{8\Gamma^2 J \beta^3\mathrm{coth}\varepsilon}{\varepsilon}$, and the dummy index $n$ in $\prod_n^{'}$ runs over all positive odd integers from one to infinity. The constant $\hat{C}=\prod_n^{'}(\frac{2\beta J}{\pi})$. Performing the gaussian integrals, and using the formula
\begin{equation}
\prod_{n=1}^{\infty}
\left(
1
+
\frac{z_1}{(2n-1)^2+z_2}
\right)
=
\frac{\cosh \frac{\pi}{2}\sqrt{z_1+z_2}}{\cosh \frac{\pi}{2}\sqrt{z_2}}
,
\label{eq.V.C.11}
\end{equation}
we obtain
\begin{equation}
Z_Q
\stackrel{\mathrm{2nd}}{\longrightarrow}
e^{-\beta N \hat{f}_s}
\frac{\cosh \varepsilon}{\cosh\sqrt{\varepsilon^2-\frac{\hat{g}}{4}}}
,
\label{eq.V.C.12}
\end{equation}
valid in the paramagnetic phase $\Gamma\ge 2J$.

\subsection{Energy gap $E_1-E_0$ in the paramagnetic phase}
\label{sec.sub.parity gap in para phase}

From Eqs. (\ref{eq.IV.A.07}) and (\ref{eq.V.C.12}), one has
\begin{align}
Z-Z_Q
&
\stackrel{\mathrm{2nd}}{\longrightarrow}
\,
e^{-\beta N f_s}
\,
\frac{\sinh\varepsilon}{\sinh\sqrt{\varepsilon^2-\frac{g}{4}}}
\left[
1
-
e^{-\beta N (\hat{f}_s-f_s)}
\,\,
\mathrm{coth}\varepsilon
\,\,
\frac{\sinh\sqrt{\varepsilon^2-\frac{g}{4}}}{\cosh\sqrt{\varepsilon^2-\frac{\hat{g}}{4}}}
\right]
\label{eq.V.D.01}\\[10pt]
&
=
e^{-\beta N f_s}
\frac{\sinh\beta \Gamma}{\sinh \beta\sqrt{\Gamma(\Gamma-2J\tanh\varepsilon)}}
\left[
1
-
\left(\tanh^{N-1}\beta\Gamma\right)
\frac{\sinh \beta\sqrt{\Gamma(\Gamma-2J\tanh\varepsilon)}}{\cosh \beta\sqrt{\Gamma(\Gamma-2J\,\mathrm{coth}\,\varepsilon)}}
\right],
\label{eq.V.D.02}
\end{align}
where in the second line we have used the fact that in the paramagnetic phase $m_s=0$ and $\varepsilon=\beta\Gamma$. When $\beta$ is large, one can approximate the $\tanh\varepsilon$ and $\mathrm{coth}\,\varepsilon$ appearing inside radicals by 1. Expanding $\tanh\beta\Gamma$ and $\tanh\beta\sqrt{\Gamma(\Gamma-2J)}$ using the series expansion $\tanh\theta=1+2\sum_{n=1}^{\infty}(-1)^n e^{-2n\theta}$, one obtains
\begin{equation}
Z-Z_Q
\stackrel{\mathrm{2nd}}{\longrightarrow}
e^{N\ln2\cosh\beta\Gamma}
\frac{\sinh\beta\Gamma}{\sinh\beta\sqrt{\Gamma(\Gamma-2J)}}
\cdot
2
e^{-2\beta\sqrt{\Gamma(\Gamma-2J)}}
\cdot
\left[
1+
O
\left(
e^{-2\beta\left(\Gamma-\sqrt{\Gamma(\Gamma-2J)}\right)}
\right)
\right].
\label{eq.V.D.03}
\end{equation}
Inserting Eq. (\ref{eq.V.D.03}) into Eq. (\ref{eq.V.A.06}), we get
\begin{equation}
E_1
\stackrel{\mathrm{2nd}}{\longrightarrow}
-N\Gamma -\Gamma + 3\sqrt{\Gamma(\Gamma-2J)}
\,\,\,\,\,\,\,\,\,\,
\mathrm{for} 
\,\,\,\,
\Gamma\ge 2J.
\label{eq.V.D.04}
\end{equation}
Subtracting away $E_0$ given by Eq. (\ref{eq.IV.A.09}), one has
\begin{equation}
E_1-E_0
\stackrel{\mathrm{2nd}}{\longrightarrow}
2\sqrt{\Gamma(\Gamma-2J)}
\,\,\,\,\,\,\,\,\,\,
\mathrm{for} 
\,\,\,\,
\Gamma\ge 2J.
\label{eq.V.D.05}
\end{equation}
Comparing with Eq. (\ref{eq.II.A.03}), we see that we have recovered the energy gap obtained by Holstein-Primakoff transform in the paramagnetic phase.

\section{Energy gap using excitation operator}
\label{sec.gap 2}

\subsection{Formulation of $E_0+E_1$}
\label{sec.sub.gap 2 formulation}

In this section, we derive the second formula to calculate the energy gap. Like Sec. \ref{sec.sub.gap 1 formulate}, the formulation in Sec. \ref{sec.sub.gap 2 formulation} is general and not specific to any model. 

Define
\begin{equation}
Z_A
=
\mathrm{Tr}
\left(
Ae^{-2\beta H}Ae^{-2\beta H}
\right),
\label{eq.VI.A.01}
\end{equation}
where $A$ is a hermitian operator. Then, for a suitable choice of $A$,
\begin{equation}
E_0+E_1
=
\lim_{\beta\rightarrow\infty}
-
\frac{1}{2\beta}
\ln Z_A.
\label{eq.VI.A.02}
\end{equation}
Before deriving Eq. (\ref{eq.VI.A.02}), we first introduce two selection rules.

\textbf{Selection rule 1}

Let $|q\rangle$ and $|q^{\prime}\rangle$ be eigenvectors of a parity operator $Q$ with eigenvalues $q$ and $q^{\prime}$, respectively. If $A$ and $Q$ anti-commute, i.e., $QA+AQ=0$, then,
\begin{equation}
\langle q|A|q^{\prime}\rangle=0
\label{eq.VI.A.03}
\end{equation}
unless $q=-q^{\prime}$; in other words, $A$ only connects states with opposite parity \cite{note.proof of selection rule of A}. 

\textbf{Selection rule 2}

For a Hamiltonian $H$, if one can find an operator $A^{\prime}$ such that
\begin{equation}
[H,A^{\prime}]=cA,
\label{eq.VI.A.04}
\end{equation}
where $c$ is a non-zero constant, then
\begin{equation}
\langle E_n^{a}|A|E_n^{b}\rangle=0.
\label{eq.VI.A.05}
\end{equation}
In other words, the matrix element of $A$ between the same or degenerate energy eigenstates vanishes \cite{note.proof of selection rule using Aprime}. 

We now derive Eq. (\ref{eq.VI.A.02}). From the expansion Eq. (\ref{eq.V.A.02}) for $e^{-\beta H}$, we have
\begin{align}
e^{-\beta H}Ae^{-\beta H}
&
=
e^{-2\beta E_0} \left( \sum_{a,b} |E_0^a\rangle\langle E_0^a|A|E_0^b\rangle\langle E_0^b| \right)
+
e^{-\beta (E_0+E_1)} \left(\sum_{a,b}|E_0^a\rangle\langle E_0^a|A|E_1^b\rangle\langle E_1^b| \right)
\nonumber\\
&
+
e^{-\beta (E_0+E_1)}\left( \sum_{a,b}|E_1^a\rangle\langle E_1^a|A|E_0^b\rangle\langle E_0^b| \right)
+
e^{-2\beta E_1}\left( \sum_{a,b}|E_1^a\rangle\langle E_1^a|A|E_1^b\rangle\langle E_1^b| \right)
+
\cdots,
\label{eq.VI.A.06}
\end{align}
where each dummy index in $\sum_{a,b}$ runs over the quantum numbers of the energy level it is being tagged with. We now eliminate the coefficients of $e^{-2\beta E_0}$ and $e^{-2\beta E_1}$ in two different ways by appealing to the two selection rules. If the two conditions of Sec. \ref{sec.sub.gap 1 formulate} hold and $A$ anti-commutes with $Q$, by selection rule 1, $\langle E_0^a|A|E_0^b\rangle=\langle E_1^a|A|E_1^b\rangle=0$, and the first and fourth terms vanish. Similarly, if one can show Eq. (\ref{eq.VI.A.04}), then these matrix elements also vanish because of selection rule 2. Note that the two selection rules are not mutually exclusive, and it is possible for both to function at the same time. 

We then have
\begin{equation}
e^{-\beta H}Ae^{-\beta H}=
e^{-\beta(E_0+E_1)}
\left(
\sum_{a,b}
\sum_{p=\pm1}
|E_{\frac{1-p}{2}}^a\rangle\langle E_{\frac{1-p}{2}}^a|
A
|E_{\frac{1+p}{2}}^b\rangle\langle E_{\frac{1+p}{2}}^b|
\right)
+
\cdots.
\label{eq.VI.A.07}
\end{equation}
Squaring both sides,
\begin{equation}
\left(
e^{-\beta H}Ae^{-\beta H}
\right)^2
=
e^{-2\beta(E_0+E_1)}
\left(
\sum_{a,b,c}
\,
\sum_{p=\pm1}
|E^a_{\frac{1-p}{2}} \rangle\langle E^a_{\frac{1-p}{2}} |A|E^b_{\frac{1+p}{2}} \rangle \langle E_{\frac{1+p}{2}}^b|A|E_{\frac{1-p}{2}}^c\rangle\langle E_{\frac{1-p}{2}}^c|
\right)
+
\cdots,
\label{eq.VI.A.08}
\end{equation}
where we have used $\langle E_n^a|E_{m}^b\rangle=\delta_{nm}\delta_{ab}$. Taking trace and using the cyclic permutation property of trace, one obtains
\begin{equation}
\mathrm{Tr}
\left(
Ae^{-2\beta H}Ae^{-2\beta H}
\right)
=
2
e^{-2\beta(E_0+E_1)}
\left(
\sum_{a,b}
|\langle E_0^a|A| E_1^b\rangle|^2
\right)
\left[
1
+
O\left(e^{-\beta(E_2-E_1)}\right)
\right].
\label{eq.VI.A.09}
\end{equation}
Assuming that $\sum_{a,b}|\langle E_0^a|A| E_1^b\rangle|^2$ does not vanish, the trace formula Eq. (\ref{eq.VI.A.02}) follows. Physically, the non-vanishing of the sum of the matrix elements means that $A$ must connect the subspaces of the two energy levels. 

The key step in our derivation lies in the method of eliminating of the leading $e^{-2\beta E_0}$ term in Eq. (\ref{eq.VI.A.06}). When appealing to selection rule 1, one essentially requires that the ground-state energy level, if degenerate, has a definite parity. This condition might be too restrictive in actual applications. On the other hand, selection rule 2 does not make any assumption about parity, and is applicable even if the ground-state energy level consists of a mixture of parity states.

We summarize the conditions used during the derivation. If one appeals to selection rule 1, one needs the following conditions:
\begin{enumerate}
\item The 2 conditions of Sec. \ref{sec.sub.gap 1 formulate}.

\item The hermitian operator $A$ anti-commutes with $Q$. 

\item $\sum_{a,b}|\langle E_0^a|A| E_1^b\rangle|^2$ is non-zero.
\end{enumerate}
If one appeals to selection rule 2, one needs the following conditions:
\begin{enumerate}
\item Existence of an operator $A^{\prime}$ such that $[H,A^{\prime}]=cA$, where the constant $c\ne 0$. 

\item $\sum_{a,b}|\langle E_0^a|A| E_1^b\rangle|^2$ is non-zero.
\end{enumerate}

\subsection{Physical intepretation of $A$ as excitation operator}
\label{sec.sub.gap 2 interpretation}

We now specialize our discussion to the ferromagnetic model. The operator
\begin{equation}
A_z
=
\frac{1}{\sqrt{N}}
\sum_{i=1}^N
\sigma^z_i
\label{eq.VI.B.01}
\end{equation}
anti-commutes with the $Q$ of Eq. (\ref{eq.V.B.01}), and is a possible candidate for $A$ by appealing to selection rule 1. To see its physical significance, consider the deep paramagnetic regime $J=0$. The ground-state $|E_0\rangle_{J=0}$ and first excited-state $|E_1\rangle_{J=0}$ (c.f. Appendix \ref{sec.appendix.ZQ in para phase}) are related by 
\begin{equation}
|E_1\rangle_{J=0}
\propto
A_z
|E_0\rangle_{J=0}.
\label{eq.VI.B.02}
\end{equation}
Hence, $A_z$ excites the ground-state to the first excited-state. The trace formula Eq. (\ref{eq.VI.A.02}), therefore, obtains information about $E_1$ by choosing a suitable $A$ that functions as an excitation operator.  

The relation Eq. (\ref{eq.VI.B.02}) holds generally in the paramagnetic phase. On the other hand, in the ferromagnetic phase $A_z$ is no longer an excitation operator; for instance, when $\Gamma=0$, the ground-state stays within its own subspace after being acted on by $A_z$. Hence, $A_z$ cannot be used in the ferromagnetic phase. 

The choice of excitation operator is not unique. Consider the operator,
\begin{equation}
A_y
=
\frac{1}{\sqrt{N}}
\sum_{i=1}^N
\sigma^y_i
.
\label{eq.VI.B.03}
\end{equation}
$A_y$ anti-commutes with $Q$, and Eq. (\ref{eq.VI.B.02}) is also valid if one replaces $A_z$ by $A_y$. However, $A_y$ is different from $A_z$ in that it is also an excitation operator in the ferromagnetic phase. For instance, when $\Gamma=0$, the ground-states $|E_0^{\pm}\rangle_{\Gamma=0}$ and first excited-states $|E_1^{\pm}\rangle_{\Gamma=0}$ (c.f. Appendix \ref{sec.appendix.ZA}) are also related by $|E_1^{\pm}\rangle_{\Gamma=0} \propto A_y |E_0^{\pm}\rangle_{\Gamma=0}$. Hence, $A_y$ can be used in both phases. 

In the above, we have seen that $A_z$ and $A_y$ act as excitation operators by looking at $J=0$ and at $\Gamma=0$. More generally when both $J$ and $\Gamma$ are non-zero, this property is quantified by the non-vanishing of $\sum_{a,b}|\langle E_0^a|A| E_1^b\rangle|^2$. In Appendix \ref{sec.appendix.ZA}, we show that this condition is indeed satisfied for $A_z$ and $A_y$. In addition, the other conditions which are needed to establish the trace formula Eq. (\ref{eq.VI.A.02}) for these two operators will also be verified there.

\subsection{Path integral representation of $Z_{A_{\mu}}$}
\label{sec.sub.gap 2 path integral representation}

Let us introduce the notation $A_{\mu}$ where $\mu$ can be $z$ or $y$. From Eq. (\ref{eq.VI.A.01}), one has
\begin{align}
Z_{A_{\mu}}
&
=
\mathrm{Tr}
\left[
\left(
\frac{1}{\sqrt{N}}
\sum_{i=1}^{N}
\sigma_i^{\mu}
\right)
e^{-2\beta H}
\left(
\frac{1}{\sqrt{N}}
\sum_{i=1}^{N}
\sigma_i^{\mu}
\right)
e^{-2\beta H}
\right]
\label{}\nonumber\\[5pt]
&
=
\frac{1}{N}
\sum_{i,j}
\mathrm{Tr}
\left(
\sigma_i^{\mu}
e^{-2\beta H}
\sigma_j^{\mu}
e^{-2\beta H}
\right)
\label{}\nonumber\\
&
=
\frac{1}{N}
\sum_{i}
\mathrm{Tr}
\left(
\sigma_i^{\mu}
e^{-2\beta H}
\sigma_i^{\mu}
e^{-2\beta H}
\right)
+
\frac{1}{N}
\sum_{i \ne j}
\mathrm{Tr}
\left(
\sigma_i^{\mu}
e^{-2\beta H}
\sigma_j^{\mu}
e^{-2\beta H}
\right)
\label{eq.VI.C.01}\\
&
=
\mathrm{Tr}
\left(
\sigma_i^{\mu}
e^{-2\beta H}
\sigma_i^{\mu}
e^{-2\beta H}
\right)
+
(N-1)
\mathrm{Tr}
\left(
\sigma_i^{\mu}
e^{-2\beta H}
\sigma_j^{\mu}
e^{-2\beta H}
\right)
\label{eq.VI.C.02}
\end{align}
In going from the third to the fourth line, we have used the fact that all spins and pairs of spins are identical in our ferromagnetic model. The calculation of the two traces in Eq. (\ref{eq.VI.C.02}) is the same as described in Sec. \ref{sec.sub.path integral representation}, i.e., applying Suzuki-Trotter decomposition to each of the two $e^{-2\beta H}$ and introducing order parameters $m_{\kappa}$ at each Trotter slice. An additional step is to factor out the spin indices involved with Pauli matrices ($i$ for the first trace, $i$ and $j$ for the second one) for separate calculation. For the first trace, the $i$th spin encounters $\sigma^{\mu}$ two times along the Trotter dimension, once at $\kappa=M-1$ and another time at $\kappa=2M-1$. For the second trace, the $i$th spin encounters $\sigma^{\mu}$ once at $\kappa=2M-1$ and the $j$th spin once at $\kappa=M-1$. For the rest of the spins indices not involved with Pauli matrices, their calculation is the same as that for $Z$. The path integral representation of $Z_{A_{\mu}}$ is then 
\begin{equation}
\left(
Z_{A_{\mu}}
\right)_M
=
\left(
\sqrt{\frac{2\beta J N}{\pi M}}
\right)^{2M}
\prod_{\kappa=0}^{2M-1}
\int_{-\infty}^{\infty}
dm_{\kappa}
\,
\frac{\mathcal{T}_0\mathcal{T}_{3\mu} + (N-1) \mathcal{T}_{1\mu}\mathcal{T}_{2\mu}}{(\mathcal{T}_0)^2}
\,
\exp
\left(
-\frac{2\beta JN}{M}\sum_{\kappa=0}^{2M-1}m_{\kappa}^2
+
N
\ln
\mathcal{T}_0
\right)
,
\label{eq.VI.C.03}
\end{equation}
where
\begin{align}
\mathcal{T}_{0}
&
=
\sum_{\sigma=\pm1}
\langle \sigma |
\left[
\prod_{\kappa=0}^{2M-1}
e^{\frac{1}{M}(2\beta\Gamma\sigma^x+4\beta J m_{\kappa}\sigma^z)}
\right]
|\sigma\rangle
.
\label{eq.VI.C.04}\\[10pt]
\mathcal{T}_{1\mu}
&
=
\sum_{\sigma=\pm1}
\langle \sigma |
\sigma^{\mu}
\left[
\prod_{\kappa=0}^{2M-1}
e^{\frac{1}{M}(2\beta\Gamma\sigma^x+4\beta J m_{\kappa}\sigma^z)}
\right]
|\sigma\rangle
.
\label{eq.VI.C.05}\\[10pt]
\mathcal{T}_{2\mu}
&
=
\sum_{\sigma=\pm1}
\langle \sigma |
\left[
\prod_{\kappa=M}^{2M-1}
e^{\frac{1}{M}(2\beta\Gamma\sigma^x+4\beta J m_{\kappa}\sigma^z)}
\right]
\sigma^{\mu}
\left[
\prod_{\kappa=0}^{M-1}
e^{\frac{1}{M}(2\beta\Gamma\sigma^x+4\beta J m_{\kappa}\sigma^z)}
\right]
|\sigma\rangle
.
\label{eq.VI.C.06}\\[10pt]
\mathcal{T}_{3\mu}
&
=
\sum_{\sigma=\pm1}
\langle \sigma |
\sigma^{\mu}
\left[
\prod_{\kappa=M}^{2M-1}
e^{\frac{1}{M}(2\beta\Gamma\sigma^x+4\beta J m_{\kappa}\sigma^z)}
\right]
\sigma^{\mu}
\left[
\prod_{\kappa=0}^{M-1}
e^{\frac{1}{M}(2\beta\Gamma\sigma^x+4\beta J m_{\kappa}\sigma^z)}
\right]
|\sigma\rangle
.
\label{eq.VI.C.07}
\end{align}
In Eq. (\ref{eq.VI.C.03}), $\mathcal{T}_{0}\mathcal{T}_{3\mu}$ and $(N-1)\mathcal{T}_{1\mu}\mathcal{T}_{2\mu}$ come from the first and second terms of Eq. (\ref{eq.VI.C.02}), respectively.

The single spin traces Eqs. (\ref{eq.VI.C.04}) to (\ref{eq.VI.C.07}) contain the familiar product sequence which can be interpreted as the fundamental matrix propagating a spinor between two time points, as described in Sec. \ref{sec.sub.ode interpretation}. $\mathcal{T}_0$ and $\mathcal{T}_{1\mu}$ have the same forms as $\mathcal{T}$ and $\mathcal{T}_Q$, respectively. For $\mathcal{T}_{2\mu}$, one propagates the initial spinor to time $\kappa=M-1$, multiplies it by $\sigma^{\mu}$, and then continue propagating it until $\kappa=2M-1$ before taking the inner product. $\mathcal{T}_{3\mu}$ is similar to $\mathcal{T}_{2\mu}$, but with an additional final step of multiplying by $\sigma^{\mu}$ before taking inner product. Eqs. (\ref{eq.VI.C.04}) to (\ref{eq.VI.C.07}) can all be calculated by following the prescription of Sec. \ref{sec.sub.expansion of T}, and  will be discussed in the next section. 

In Eq. (\ref{eq.VI.C.03}), the second (exponential) integrand is, apart from a rescaling of the constants $J$, $\Gamma$, and $M$, the same as that appearing in $Z$. From Eq. (\ref{eq.IV.A.05}), we see that at second-order approximation, this term becomes a gaussian function. On the other hand, the first integrand will turn out to be quadratic after making the appropriate approximations. Hence, the path integral Eq. (\ref{eq.VI.C.03}) is of the form $\int x^2 e^{-x^2}dx$ and can be integrated easily. 

\subsection{Perturbative expansions of the single spin traces $\mathcal{T}_{0}$, $\mathcal{T}_{1\mu}$, $\mathcal{T}_{2\mu}$, and $\mathcal{T}_{3\mu}$}
\label{sec.sub.gap 2 expand Ts}

The perturbative expansion of $\mathcal{T}_0$ can be obtained from previous results of $\mathcal{T}$ simply by extending the upper integration limit in Eq. (\ref{eq.III.C.11}) from 1 to 2 and making the substitutions $J\rightarrow 2J$ and $\Gamma\rightarrow 2\Gamma$ in Eqs. (\ref{eq.III.C.13}) to (\ref{eq.III.C.17}) \cite{note.explanation for scaling of two in J G and integration limits}. The $\mathcal{T}_0$ appearing in the exponent of the second integrand of Eq. (\ref{eq.VI.C.03}) is expanded to second order as before. However, it is not necessary to do that for every term of the first integrand $\frac{\mathcal{T}_{0}\mathcal{T}_{3\mu} +(N-1) \mathcal{T}_{1\mu}\mathcal{T}_{2\mu}}{(\mathcal{T}_0)^2}$. The leading non-vanishing term in the perturbative expansions of $\mathcal{T}_{1\mu}$, $\mathcal{T}_{2\mu}$, and $\mathcal{T}_{3\mu}$ is of order $\lambda^1$, $\lambda^1$, and $\lambda^0$, respectively. If we keep just these leading terms and the static approximation of $\mathcal{T}_0$, we will obtain the leading order-$N^0$ term of the first integrand, which is sufficient for our subsequent calculations.

The perturbative expansions of the traces Eqs. (\ref{eq.VI.C.04}) to (\ref{eq.VI.C.07}) can again be calculated by following the steps of Sec. \ref{sec.sub.expansion of T}. For $\mathcal{T}_{2\mu}$ and $\mathcal{T}_{3\mu}$, the time evolution is interrupted halfway by a Pauli matrix, and the integration limits in the integrals of $m_d(t)$ are affected. Let us introduce the notation
\begin{equation}
M_s^{t_1,t_2}
=
\int_{t_1}^{t_2}
dt
\,
m_d(t)
\,
e^{s4\varepsilon t}
,
\label{eq.VI.D.01}
\end{equation}
where the subscript $s$ has the same meaning as in Eq. (\ref{eq.III.C.11}), and $\varepsilon=\sqrt{(\beta\Gamma)^2+(2\beta J m_s)^2}$. Noting that the Pauli matrices take the form $\sigma^z={\alpha\,\,\gamma\choose\gamma\,\,-\alpha}$ and $\sigma^y={0\,\,-i\choose i\,\,0}$ in the basis where $2\mathcal{H}_s$ is diagonal, we have for $\mu=y$, 
\begin{align}
\mathcal{T}_{1y}
&
\stackrel{\mathrm{1st}}{\longrightarrow}
i \lambda (4\beta J)  \gamma
(
e^{4\varepsilon} M_{-}^{0,2} 
-
e^{-4\varepsilon} M_{+}^{0,2} 
).
\label{eq.VI.D.02}\\
\mathcal{T}_{2y}
&
\stackrel{\mathrm{1st}}{\longrightarrow}
i \lambda (4\beta J) \gamma
(
M_-^{0,1}
-
M_+^{0,1}
+
e^{8\varepsilon}
M_-^{1,2}
-
e^{-8\varepsilon}
M_+^{1,2}
).
\label{eq.VI.D.03}\\
\mathcal{T}_{3y}
&
\stackrel{\mathrm{s.a.}}{\longrightarrow}
2.
\label{eq.VI.D.04}
\end{align}
The notation $\stackrel{\mathrm{1st}}{\longrightarrow}$ means first-order approximation. For $\mu=z$, we have
\begin{align}
\mathcal{T}_{1z}
&
\stackrel{\mathrm{1st}}{\longrightarrow}
2\alpha\sinh 4\varepsilon
+
\lambda
(4\beta J)
[
2
\alpha^2
\cosh 4\varepsilon
\,
M_0^{0,2}
+
\gamma^2
(
e^{4\varepsilon}
M_-^{0,2}
+
e^{-4\varepsilon}
M_+^{0,2}
)
].
\label{eq.VI.D.05}\\
\mathcal{T}_{2z}
&
\stackrel{\mathrm{1st}}{\longrightarrow}
2\alpha\sinh4\varepsilon
+
\lambda(4\beta J)
[
2\alpha^2\cosh 4\varepsilon
\,
M_0^{0,2}
+
\gamma^2
(
M_+^{0,1}
+
M_-^{0,1}
+
e^{8\varepsilon}
M_-^{1,2}
+
e^{-8\varepsilon}
M_+^{1,2}
)
].
\label{eq.VI.D.06}\\
\mathcal{T}_{3z}
&
\stackrel{\mathrm{s.a.}}{\longrightarrow}
2
(
\alpha^2 \cosh 4\varepsilon
+
\gamma^2
).
\label{eq.VI.D.07}
\end{align}
For completeness, we note that $\mathcal{T}_0\stackrel{\mathrm{s.a.}}{\longrightarrow} 2\cosh 4\varepsilon$. As an example of one of these calculations, the derivation of Eq. (\ref{eq.VI.D.03}) for $\mathcal{T}_{2y}$ is given in Appendix \ref{sec.appendix.calculating T2y}.

\subsection{Calculation of $Z_{A_y}$}
\label{sec.sub.gap 2 calculate ZA}

The path integral Eq. (\ref{eq.VI.C.03}) is again performed by making the non-static ansatz Eq. (\ref{eq.III.B.01}) and expanding $m_d(t)$ in Fourier series. Due to the presence of two $e^{-2\beta H}$ in $Z_{A_{\mu}}$, the length of each path is doubled. The Fourier expansion respecting this boundary condition is
\begin{equation}
m_d(t)
=
\sum_{n=-\infty}^{\infty}
c_n
\,
e^{i\pi n t}
,
\label{eq.VI.E.01}
\end{equation}
where $0<t<2$.

We first consider the exponential integrand. Repeating the derivation of Eq. (\ref{eq.IV.A.05}) with the changes $J\rightarrow 2J$, $\Gamma\rightarrow 2\Gamma$, $M\rightarrow 2M$, and with the expansion Eq. (\ref{eq.VI.E.01}), we have
\begin{equation}
Z_{A_{\mu}}
\stackrel{\mathrm{2nd}}{\longrightarrow}
C'
e^{-\beta N f'_s}
\int
dc_0
\prod_{n=1}^{\infty}
dc_n dc_n^*
\,
\frac{\mathcal{T}_0 \mathcal{T}_{3\mu} + (N-1)\mathcal{T}_{1\mu} \mathcal{T}_{2\mu} }{(\mathcal{T}_0)^2}
\,
\exp
\left(
-
4\beta J
\sum_{n=-\infty}^{\infty}
g'_n c_n c_{-n}
\right)
,
\label{eq.VI.E.02}
\end{equation}
where $f'_s=4Jm_s^2-\frac{1}{\beta}\ln 2 \cosh 4\varepsilon$, $g'_n=1-\frac{g'}{(\pi n)^2+(4\varepsilon)^2}$ with $g'=\frac{32\Gamma^2 J \beta^3 \tanh4\varepsilon}{\varepsilon}$, and the constant $C'=\sqrt{\frac{4\beta J}{\pi}}\prod_{n=1}^{\infty}(\frac{8\beta J}{\pi})$. 

We now consider the second non-gaussian integrand, for the case of $\mu=y$. Inserting the expansion Eq. (\ref{eq.VI.E.01}) into Eqs. (\ref{eq.VI.D.02}) and (\ref{eq.VI.D.03}), we have 
\begin{equation}
\frac{\mathcal{T}_0 \mathcal{T}_{3y} + (N-1)\mathcal{T}_{1y} \mathcal{T}_{2y} }{(\mathcal{T}_0)^2}
=
\mathrm{sech}4\varepsilon
-
\left(\frac{g'}{4\beta\Gamma}\right)^2
\sum_{n=-\infty}^{\infty}
c_n c_{-n}
\frac{(-1)^n (\pi n)^2}{[(\pi n)^2+(4\varepsilon)^2]^2}
+
O(N^{-1/2})
,
\label{eq.VI.E.03}
\end{equation}
where we have dropped the cross terms $c_n c_m$ because they will vanish after integrating over by the gaussian function. Performing the gaussian integrals \cite{note. gaussian formula for second moment}, we have 
\begin{equation}
Z_{A_y}
\stackrel{O(1)}{\longrightarrow}
e^{-\beta N f'_s}
\frac{\sinh 4\varepsilon}{\sinh 4\sqrt{\varepsilon^2-\frac{g'}{16}}}
\left(
\mathrm{sech}4\varepsilon
-
\frac{(g')^2}{64 \Gamma^2 J\beta^3}
\sum_{n=1}^{\infty}
\frac{(-1)^n(\pi n)^2}{[(\pi n)^2+(4\varepsilon)^2][(\pi n)^2+(4\varepsilon)^2-g']}
\right)
,
\label{eq.VI.E.04}
\end{equation}
where $\stackrel{O(1)}{\longrightarrow}$ denotes the combined approximations made in Eqs. (\ref{eq.VI.E.02}) and (\ref{eq.VI.E.03}) such that overall result is accurate up to the term $N^0$. Simplifying the summand of the series using partial fractions $\frac{1}{x(x-x_0)}=\frac{1}{x_0}(\frac{1}{x-x_0}-\frac{1}{x})$, and using the formula
\begin{equation}
\sum_{n=1}^{\infty}
\frac{(-1)^n}{z_1 n^2+z_2}
=
-\frac{1}{2z_2}
+
\frac{\pi}{2\sqrt{z_1 z_2}}
\mathrm{cosech}
\left(
\pi\sqrt{\frac{z_2}{z_1}}
\right)
,
\label{eq.VI.E.05}
\end{equation}
we obtain
\begin{equation}
Z_{A_y}
\stackrel{O(1)}{\longrightarrow}
e^{-\beta N f'_s}
\,
\frac{\sinh4\varepsilon \tanh4\varepsilon}{\sinh^2 4\sqrt{\varepsilon^2-\frac{g'}{16}}}
\,
\frac{\sqrt{\varepsilon^2-\frac{g'}{16}}}{\varepsilon}
.
\label{eq.VI.E.06}
\end{equation}
The result Eq. (\ref{eq.VI.E.06}) holds everywhere except at the critical point $\Gamma=2J$ where the factor $\sqrt{\varepsilon^2-\frac{g'}{16}}$ becomes zero. Indeed, at the critical point the ground and first-excited states collide and the relation Eq. (\ref{eq.VI.A.02}) is no longer valid. 

We have presented the calculation of $Z_{A_y}$. The case of $Z_{A_z}$ is similar and is given in Appendix \ref{sec.appendix.calculation of ZAz}.

\subsection{Energy gap $E_1-E_0$ in both phases}
\label{sec.sub.gap 2 gap in the 2 phases}

Inserting Eq. (\ref{eq.VI.E.06}) into Eq. (\ref{eq.VI.A.02}), we have 
\begin{equation}
E_0+E_1
\stackrel{O(1)}{\longrightarrow}
\frac{N}{2}
f'_s
+
4
\sqrt{
\Gamma^2
+
(2Jm_s)^2
-
\frac{2\Gamma^2 J}{\sqrt{\Gamma^2 + (2Jm_s)^2}}
}
-2
\sqrt{\Gamma^2 + (2Jm_s)^2}
.
\label{eq.VI.F.01}
\end{equation}
The solution of the stationary condition $\partial f'_s/\partial m_s=0$ is again given by Eq. (\ref{eq.II.C.06}). Inserting it into Eq. (\ref{eq.VI.F.01}), and subtracting away $2E_0$ with $E_0$ given by Eq. (\ref{eq.IV.A.09}), we obtain 
\begin{equation}
E_1-E_0
\stackrel{O(1)}{\longrightarrow}
\left\{
\begin{array}{ccc}
2\sqrt{\Gamma(\Gamma-2J)}  & \mathrm{for} & \Gamma \ge2J, \\
2\sqrt{(2J)^2-\Gamma^2}    & \mathrm{for} & \Gamma <2J. \\
\end{array}
\right.
\label{eq.VI.F.02}
\end{equation}
Comparing with Eq. (\ref{eq.II.A.03}), we see that we have recovered the energy gap obtained by the Holstein-Primakoff transform.

\section{Summary and discussions}
\label{sec.discussions}

In this paper, we introduced a theoretical framework for incorporating non-staticity into the path integral calculation of the partition function of quantum spin systems, thereby going beyond the static approximation. Our key observation is that the single spin trace that appears frequently in these path integrals actually evolves in time according to the Pauli equation. This re-interpretation of the trace term prompts us to solve for its time-dependent behavior by first solving the Pauli equation. Time-dependent perturbation theory is used to obtain a perturbative expansion of the solution of the Pauli equation and subsequently of the single spin trace. Upon substituting the latter into the path integral, one can then systematically integrate out the non-static component of the paths in the same manner as in conventional treatments of the Feynman kernel. 

We next applied the formalism to calculate two non-extensive quantities of an ordered spin model, the infinite-range ferromagnetic Ising model in a transverse field. We first computed the $N^0$ and $N^{-1}$ terms of the ground-state energy. For the $N^0$ term, our non-static approach reproduced the same results as that obtained using Holstein-Primakoff transform. For the $N^{-1}$ term, we checked our results by comparing with numerical calculations. The second non-extensive quantity we calculated is the energy gap between the ground and first-excited states. Two different generalized partition functions for calculating the energy of the first-excited state were proposed. The two generalized functions are cast in the form of path integrals, and the non-static method used to evaluate them. Once again, the non-static approach reproduced the results of Holstein-Primakoff transform. 

The results of Sec. \ref{sec.sub.E0 4th order} on the $N^{-1}$ term of the ground-state energy reveal a subtle point concerning what it means to expand the ground-state energy in a power series of $N^{-1}$. The curve of $N=30$ in Fig. \ref{fig.inverse N term of E0}(a) shows that the energy of a `small' system is actually not appreciably improved by the $N^{-1}$ term. The main effect of this term, rather, is to improve upon the energies of large systems in the vicinity of the critical point. 

As mentioned in the introduction, the methods for calculating the energy gap presented in this paper are also applicable to disordered models. For concreteness, let us consider the Sherrington-Kirkpatrick model in a transverse field\cite{Thirumalai89,Takahashi07,Rozenberg98,Koh16},
\begin{equation}
H_{SK}
=
-
\sum_{i=1}^N
\sum_{j>i}^N
J_{ij}
\sigma_i^z
\sigma_j^z
-
\Gamma
\sum_{i=1}^N
\sigma_i^x
,
\label{eq.VII.Hsk definition}
\end{equation}
where the couplings $J_{ij}$ are independent identical random variables drawn from a gaussian distribution with zero mean and variance $1/N$. $H_{SK}$ commutes with the parity operator $Q$ given by Eq. (\ref{eq.V.B.01}). When the couplings $J_{ij}$ are turned on from zero, the first excited-state that splits away from the original degenerate level has odd parity (c.f. Appendix \ref{sec.appendix.ZQ in para phase}). In the paramagnetic phase, the two conditions for using Eq. (\ref{eq.V.A.06}) are therefore satisfied\cite{note.concerning no collision between upper energy levels}.

The excitation operator method of Sec. \ref{sec.gap 2} also works. Let us first consider selection rule 1 and restrict ourselves to the paramagnetic phase. The operator
\begin{equation}
B_{\mu}
=
\sum_{i=1}^{N} b_i \sigma_i^{\mu}
,
\label{eq.VII.B_mu}
\end{equation}
where $\mu=z$ or $y$ and $b_i$ $(i=1,\cdots,N)$ are real parameters, anti-commutes with $Q$. We first use first-order perturbation theory to identify the first excited-state when the couplings $J_{ij}$ are turned on from zero. One then sees that by letting $(b_1,\cdots,b_N)^{\mathrm{T}}$ be the normalized eigenvector corresponding to the \emph{largest} eigenvalue of the coupling matrix $J_{ij}$, $B_{\mu}$ connects the ground-state to the first excited-state and is an excitation operator. This way of choosing $b_i$ also allows each $B_{\mu}$ to cater to the excitation of each specific realization of coupling matrix $J_{ij}$. For selection rule 2, it is easily shown that $[H_{SK},B_z]=2i\Gamma B_y$, so the conditions for using $B_y$ as an excitation operator is once again satisfied in the paramagnetic phase. 

The situation is more complicated in the spin-glass phase. Firstly, let us just consider the classical Sherrington-Kirkpatrick term without the transverse field. Different realizations of the coupling matrix $J_{ij}$ require flipping a different number of spins in order to excite the ground-state into the first excited-state. One spin-flip operators such as $B_{y}$ alone are therefore inadequate to describe the different possible modes of excitation. One needs to consider multiple spin-flip operators (e.g., a two spin-flip operator has the form $\sum_{i,j} b_{ij}\sigma_i^y\sigma_j^y$) and, in addition, know which operator to use for each realization of $J_{ij}$ to be able to calculate the gap correctly. Secondly, when $\Gamma$ is turned on, the situation is further complicated by the possibility of level crossings. If the first excited-state collides and switches places with some higher-energy levels, the number of spins needed to excite the ground-state into the new first excited-state may change. Lastly, Liu et al.\cite{Liu15} recently commented that disordered spin systems might actually be gapless in the spin-glass phase. Extremely small gaps in the spin-glass phase of Eq. (\ref{eq.VII.Hsk definition}) was also observed numerically in a recent work\cite{Koh16}. These issues---multiple-spin excitations, complications due to level crossings, and the small magnitude---highlight the difficulties of calculating the gap in the spin-glass phase of disordered systems. 

The points raised in the three preceding paragraphs will be explored more fully in our second paper.

This paper considered an Ising model, and the excitation operator $A$ is constructed using the operators $\sigma_i^{y}$ or $\sigma_i^{z}$ that flip individual Ising spins. Ising spins are, however, a bit special in that excitation is brought about by flipping. For more general types of spin elements, one should use raising or lowering operators to excite the spins. In the phase one is considering, if the $i$th spin points along, say, the $z$-direction, use $S_i^{+}=S_i^x + iS_i^y$ (or $S_i^-$) in the construction of $A$ to excite that particular spin.  

In our ferromagnetic model, one can tell simply by inspecting the Hamiltonian that the spins point along the $x$-direction in the paramagnetic phase and along the $z$-direction in the ferromagnetic phase. If we know the direction of the spins in the respective phases, we can easily construct the $A$ for each phase simply by following the instructions given in the preceding paragraph. However, it might sometimes be difficult to tell the alignment of the spins just by looking at the Hamiltonian. As an example, consider the Lipkin-Meshkov-Glick model\cite{Dusuel04,Dusuel05}
\begin{equation}
H_{LMG}
=
-\frac{h_1}{N}
\sum_{i<j}
(
\sigma^x_i\sigma^x_j
+
h_2
\sigma^y_i\sigma^y_j
)
-
h_3
\sum_i
\sigma^z_i
,
\label{eq.VII.H_LMG}
\end{equation}
where $h_1, h_2, $ and $h_3$ are parameters of the model. All three Pauli matrices $\sigma^x,\sigma^y,$ and $\sigma^z$ are involved in $H_{LMG}$, and it is difficult to tell the direction of the spins for different parameter values. In such cases, it helps to first perform a semiclassical analysis\cite{Dusuel05} to determine the average magnetization in all three directions $\langle\sum_i \sigma_i^x\rangle$, $\langle\sum_i \sigma_i^y\rangle$, and $\langle\sum_i \sigma_i^z\rangle$. This tells us the alignment of the spins. We then rotate the coordinate axes to let, say, the $z$-axis coincide with the direction of the average magnetization, after which the construction of $A$ can proceed as before in the rotated coordinate system. 

\begin{figure}[h]
\begin{center}
\includegraphics[scale=0.75]{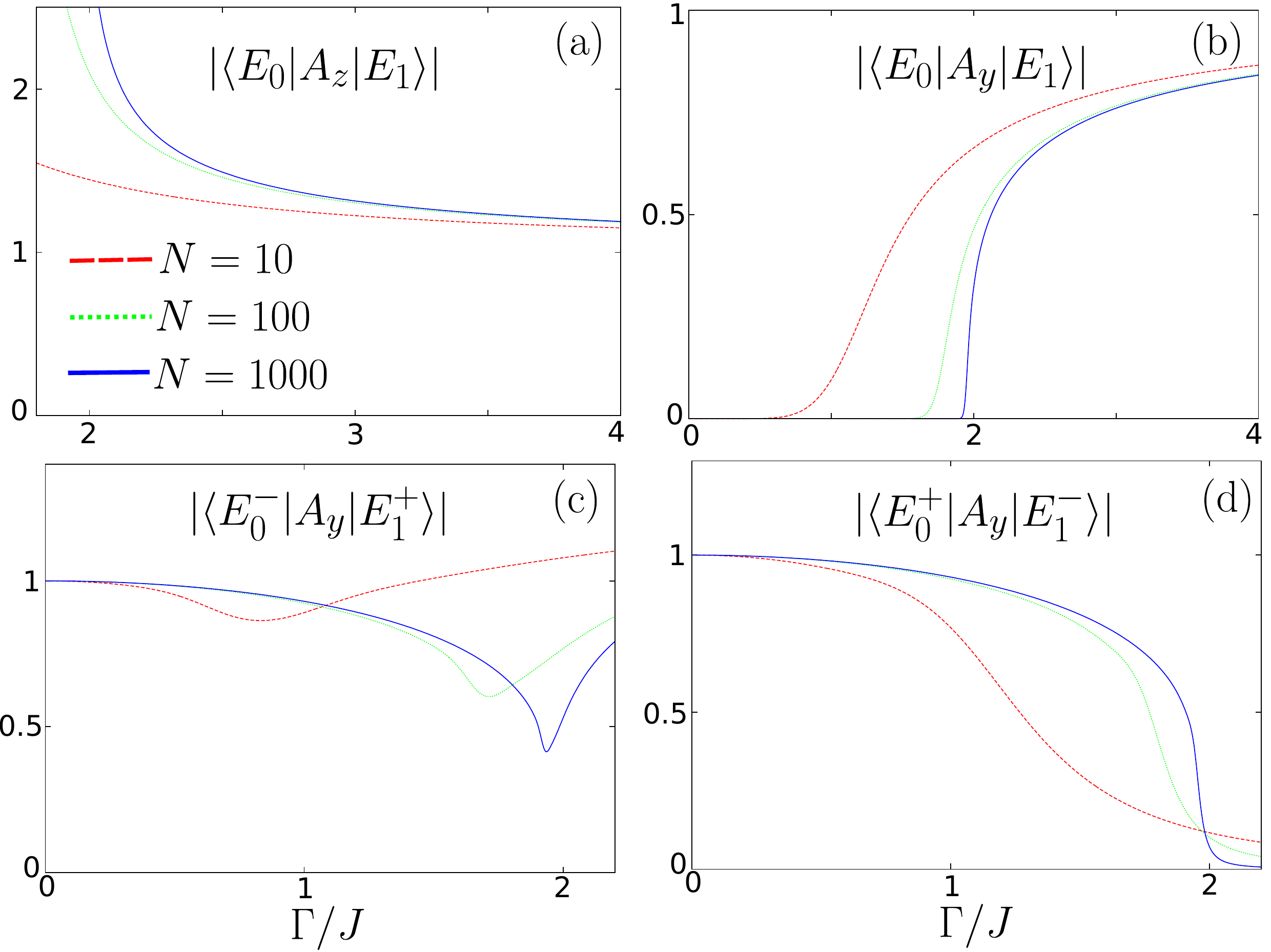}
\caption{Graphs showing the non-vanishing of $\sum_{a,b}|\langle E_0^a|A_{\mu}|E_1^b\rangle|^2$ ($\mu=z,y$) for the ferromagnetic model. Energy eigenstates are obtained by numerical diagonalization of Eq. (\ref{eq.II.A.01}) in the sector with total angular momentum $N/2$. Parity eigenstates, when indicated, are obtained by diagonalizing the parity operator. The absolute values of the matrix elements are plotted, for $N=10, 100$, and 1000. (a) For $\langle E_0|A_z|E_1\rangle$, in the paramagnetic regime. (b) For $\langle E_0|A_y|E_1\rangle$. In the ferromagnetic regime ($\Gamma<2J$), the matrix element vanishes because it becomes $\langle E_0^+|A_y|E_0^-\rangle$. (c) For $\langle E_0^-|A_y|E_1^+\rangle$, in the ferromagnetic regime.  (d) For $\langle E_0^+|A_y|E_1^-\rangle$, in the ferromagnetic regime.}
\label{fig.concerning nonvanishing of matrix elements}
\end{center}
\end{figure}

Lastly, we comment on the factor $\varepsilon^{-1}\sqrt{\varepsilon^2-(g'/16)}$ appearing in Eq. (\ref{eq.VI.E.06}). The origin of this factor should be the sum $\sum_{a,b}|\langle E_0^a|A|E_1^b\rangle|^2$ of Eq. (\ref{eq.VI.A.09}). We mentioned, after Eq. (\ref{eq.VI.E.06}), that this factor vanishes at the critical point. Indeed, this is consistent with the behavior of the matrix elements. We computed the matrix elements numerically and the results are shown in Fig. \ref{fig.concerning nonvanishing of matrix elements}. Panels (b), (c), and (d) show that for large $N$, the elements approach zero near the critical point $\Gamma=2J$. Similarly, the matrix element shown in Panel (a) diverges near the critical point, which corresponds to the singular behavior of the factor $\varepsilon\left(\sqrt{\varepsilon^2-(g'/16)}\right)^{-1}$ in Eq. (\ref{eq.calculation of ZAz.02}).

\begin{acknowledgements}
This work was partly supported by the Biomedical Research Council of A*STAR (Agency for Science, Technology and Research), Singapore. The author thanks Prof. Kazutaka Takahashi for helpful discussions. 
\end{acknowledgements}

\appendix

\section{Fourier expansions of $\mathcal{T}^{(3)}$ and $\mathcal{T}^{(4)}$}
\label{sec.appendix.fourier expansions of T3 and T4}

\subsection{Full expansion of $\mathcal{T}^{(3)}$}
\label{sec.appendix.full of T3}

$\mathcal{T}^{(3)}$ is given by Eq. (\ref{eq.III.C.16}). We need to compute Fourier expansions of the form
\begin{equation}
M_{s_1s_2s_3}
=
\sum_{n_1,n_2,n_3=-\infty}^{\infty}
c_{n_1}c_{n_2}c_{n_3}
\int_0^1 dt_1
\,
e^{(i2\pi n_1 + s_12\varepsilon) t_1}
\int_0^{t_1} dt_2
\,
e^{(i2\pi n_2 + s_22\varepsilon) t_2}
\int_0^{t_2} dt_3
\,
e^{(i2\pi n_3 + s_32\varepsilon) t_3}
.
\label{eq.fourier expansion of T3.01}
\end{equation}
The expansion of $M_{s_1s_2s_3}$ must be computed in full because one needs to take the square of $\mathcal{T}^{(3)}$ in $\frac{1}{2}(V_3)^2$. To avoid division by zero when encountering zero modes $c_0$ in the three-fold integrals, we calculated the integrals associated with the following terms individually: $c_0c_0c_0$ (3 zero modes), $c_{0}c_{0}c_{n_3},\,c_{0}c_{n_2}c_{0},\,c_{n_1}c_{0}c_{0}$ (2 zero modes), $c_{0}c_{n_2}c_{n_3},\,c_{n_1}c_{0}c_{n_3},\,c_{n_1}c_{n_2}c_{0}$ (1 zero mode), and $c_{n_1}c_{n_2}c_{n_3}$ (no zero modes). The complete triple summation is given by the sum of all these partial sums. The results are: 
  
\begin{equation}
M_{000}
=
\frac{(c_0)^3}{6}
-
\sum_{\substack{n=-\infty \\ (\ne 0)}}^{\infty}
\sum_{\substack{m=-\infty \\ (\ne 0, -n) }}^{\infty}
\frac{c_{-(n+m)}c_n c_m}{(2\pi)^2 m(n+m)}
.
\label{eq.fourier expansion of T3.02}
\end{equation}
\begin{align}
&
e^{\varepsilon}(M_{0-+}-M_{-0+}+M_{-+0}) - e^{-\varepsilon}(M_{0+-}-M_{+0-}+M_{+-0})=
\nonumber \\
&
4\left( \varepsilon \cosh\varepsilon -\sinh\varepsilon \right) \sum_{n=-\infty}^{\infty}\frac{c_0 \,c_n c_{-n}}{[n]}
-
16(2\varepsilon)^2\sinh\varepsilon\sum_{n=1}^{\infty}\frac{c_0 \,c_n c_{-n}}{[n]^2}
+
\sum_{\substack{ n=-\infty\\(\ne 0) }}^{\infty}\sum_{\substack{ m=-\infty\\(\ne 0 , -n) }}^{\infty}
c_{-(n+m)}c_n c_m \lambda_{nm} \, ,
\label{eq.fourier expansion of T3.03}
\end{align}
where
\begin{equation}
\lambda_{nm}
=
\frac{2\cosh \varepsilon}{[m][n+m]}
\left[
\frac{(2\varepsilon)^3}{i2\pi m }
+
\frac{(2\varepsilon)^3}{i2\pi(n+m)}
-
\tanh\varepsilon
\left(
(2\pi)^2 m(n+m)
+
(2\varepsilon)^2
\left(
1 + \frac{m}{n+m} + \frac{n+m}{m} 
\right)
\right)
\right]
,
\label{eq.fourier expansion of T3.04}
\end{equation}
and $[n]=(2\pi n)^2+(2\varepsilon)^2$.

\subsection{Expansion of $\mathcal{T}^{(4)}$ keeping only non-vanishing terms}
\label{sec.appendix.non-vanishing of T4}

$\mathcal{T}^{(4)}$ is given by Eq. (\ref{eq.III.C.17}). The expansions of $M_{s_1s_2s_3s_4}$ are quadruple summations. For each $M_{s_1s_2s_3s_4}$, we only need to calculate integrals associated with those $c_{n_1}c_{n_2}c_{n_3}c_{n_4}$ that do not vanish upon integration by the gaussian in Eq. (\ref{eq.IV.B.02}). There are 14 such non-vanishing terms: $c_{0}c_{0}c_{0}c_{0}$, $c_{0}c_{0}c_{n}c_{-n}$, $c_{0}c_{n}c_{0}c_{-n}$, $c_{0}c_{n}c_{-n}c_{0}$, $c_{n}c_{0}c_{0}c_{-n}$, $c_{n}c_{0}c_{-n}c_{0}$, $c_{n}c_{-n}c_{0}c_{0}$, $c_{n}c_{-n}c_{n}c_{-n}$, $c_{n}c_{-n}c_{-n}c_{n}$, $c_{n}c_{n}c_{-n}c_{-n}$, $c_{n}c_{-n}c_{m}c_{-m}$, $c_{n}c_{m}c_{-n}c_{-m}$, $c_{n}c_{m}c_{-m}c_{-n}$, and $c_{n}c_{n}c_{n}c_{n}$\cite{note.concerning vanishing of cncncncn term upon summation}. Note that the ordering of the subscripts is important for writing down the associated integral (e.g. the integrals for $c_{0}c_{0}c_{n}c_{-n}$ and $c_{0}c_{n}c_{0}c_{-n}$ are different). 

For $M_{0000}$, the expansion is
\begin{equation}
M_{0000}
\stackrel{\mathrm{n.v.}}{=}
\frac{(c_0)^4}{24}
.
\label{eq.fourier expansion of T4.01}
\end{equation}
where $\stackrel{\mathrm{n.v.}}{=}$ denotes `the non-vanishing terms'.

For $e^{\varepsilon}M_{-+-+}+e^{-\varepsilon}M_{+-+-}$, the expansion is
\begin{align}
&
e^{\varepsilon}M_{-+-+}+e^{-\varepsilon}M_{+-+-} 
\stackrel{\mathrm{n.v.}}{=}
\nonumber \\
&
\frac{(c_0)^4}{(2\varepsilon)^2}
\left(
\cosh\varepsilon - \frac{\sinh\varepsilon}{\varepsilon}
\right)
+
(c_0)^2
\sum_{\substack{n=-\infty\\(\ne 0)}}^{\infty}
c_nc_{-n}
\left[
\frac{2(\cosh\varepsilon - \frac{\sinh\varepsilon}{\varepsilon})}{[n]}
-
\frac{8\varepsilon\sinh\varepsilon}{[n]^2}
\right]
\nonumber \\
&
+
\sum_{\substack{n=-\infty\\(\ne 0)}}^{\infty}
(c_nc_{-n})^2
\left[
\frac{ 2(2\varepsilon)^2 \cosh\varepsilon + 4\varepsilon\sinh\varepsilon   }{[n]^2}
-
\frac{8(2\varepsilon)^3\sinh\varepsilon}{[n]^3}
\right]
\nonumber \\
&
+
\sum_{\substack{ n=-\infty\\ (\ne 0) }}^{\infty}
\sum_{\substack{ m=-\infty\\ (\ne 0,n,-n) }}^{\infty}
c_nc_{-n}c_mc_{-m}
\left[
\frac{(2\varepsilon)^2\cosh\varepsilon}{[n][m]}
-
4\varepsilon\sinh\varepsilon
\frac{(2\varepsilon)^2 [(2\pi n)^2+(2\pi m)^2+3(2\varepsilon)^2] - (2\pi n)^2(2\pi m)^2  }{[n]^2[m]^2}
\right]
.
\label{eq.fourier expansion of T4.02}
\end{align}

Let us denote
\begin{eqnarray}
\zeta
&=&
e^{\varepsilon}(M_{-+00}-M_{-0+0}+M_{0-+0}-M_{0-0+}+M_{00-+}+M_{-00+}) +
\nonumber\\
&&
e^{-\varepsilon}(M_{+-00}-M_{+0-0}+M_{0+-0}-M_{0+0-}+M_{00+-}+M_{+00-})
.
\label{eq.fourier expansion of T4.03}
\end{eqnarray}
The expansion of $\zeta$ is
\begin{equation}
\zeta
\stackrel{\mathrm{n.v.}}{=}
\zeta_{0}
+
\zeta_{n}
+
\zeta_{nn}
+
\zeta_{nm}
,
\label{eq.fourier expansion of T4.04}
\end{equation}
where
\begin{align}
\zeta_{0}
=
&
\,
(c_0)^4
\left(
\frac{\sinh\varepsilon}{2\varepsilon}
\right)
\left[
1
+
\frac{2}{\varepsilon^2}
-
\frac{2\,\mathrm{coth}\,\varepsilon}{\varepsilon}
\right]
,
\label{eq.fourier expansion of T4.05}\\
\zeta_{n}
=
&
\,
2\varepsilon\sinh\varepsilon
(c_0)^2
\sum_{\substack{n=-\infty\\(\ne 0)}}^{\infty}
c_nc_{-n}
\left[
\frac{1 + \frac{2}{\varepsilon^2} - \frac{2\coth\varepsilon}{\varepsilon}  }{[n]}
+
\frac{8-16\varepsilon\coth\varepsilon}{[n]^2}
+
\frac{128\varepsilon^2}{[n]^3}
\right]
,
\label{eq.fourier expansion of T4.06}\\
\zeta_{nn}
=
&
\,
48\varepsilon\sinh\varepsilon
\sum_{\substack{n=-\infty \\ (\ne 0)} }^{\infty}
\frac{(c_nc_{-n})^2}{[n][2n]}
,
\label{eq.fourier expansion of T4.07}\\
\zeta_{nm}
=
&
\,
4\varepsilon\sinh\varepsilon
\sum_{\substack{n=-\infty \\(\ne0) }}^{\infty}
\sum_{\substack{m=-\infty \\(\ne 0,n,-n) }}^{\infty}
c_nc_{-n}c_mc_{-m}\times
\nonumber\\
&
\,
\left[
\frac{7+2\frac{m}{n}+6\frac{n}{m}}{[n][n+m]}
-
\frac{1}{(2\pi)^2nm[n+m]}
+
\frac{2(2\varepsilon)^2-(2\pi)^2nm}{[n][m][n+m]}
+
\frac{(2\varepsilon)^2-(2\pi n)^2-2(2\pi)^2n(n+m)}{[n]^2[n+m]}
\right]
.
\label{eq.fourier expansion of T4.08}
\end{align}

\section{Fourier expansions of terms in $\left(\mathcal{T}^{(3)}\right)^2$ which are product of double summations}
\label{sec.appendix.calculation of half V3 square}

\begin{figure}[h]
\begin{center}
\includegraphics[scale=1.1]{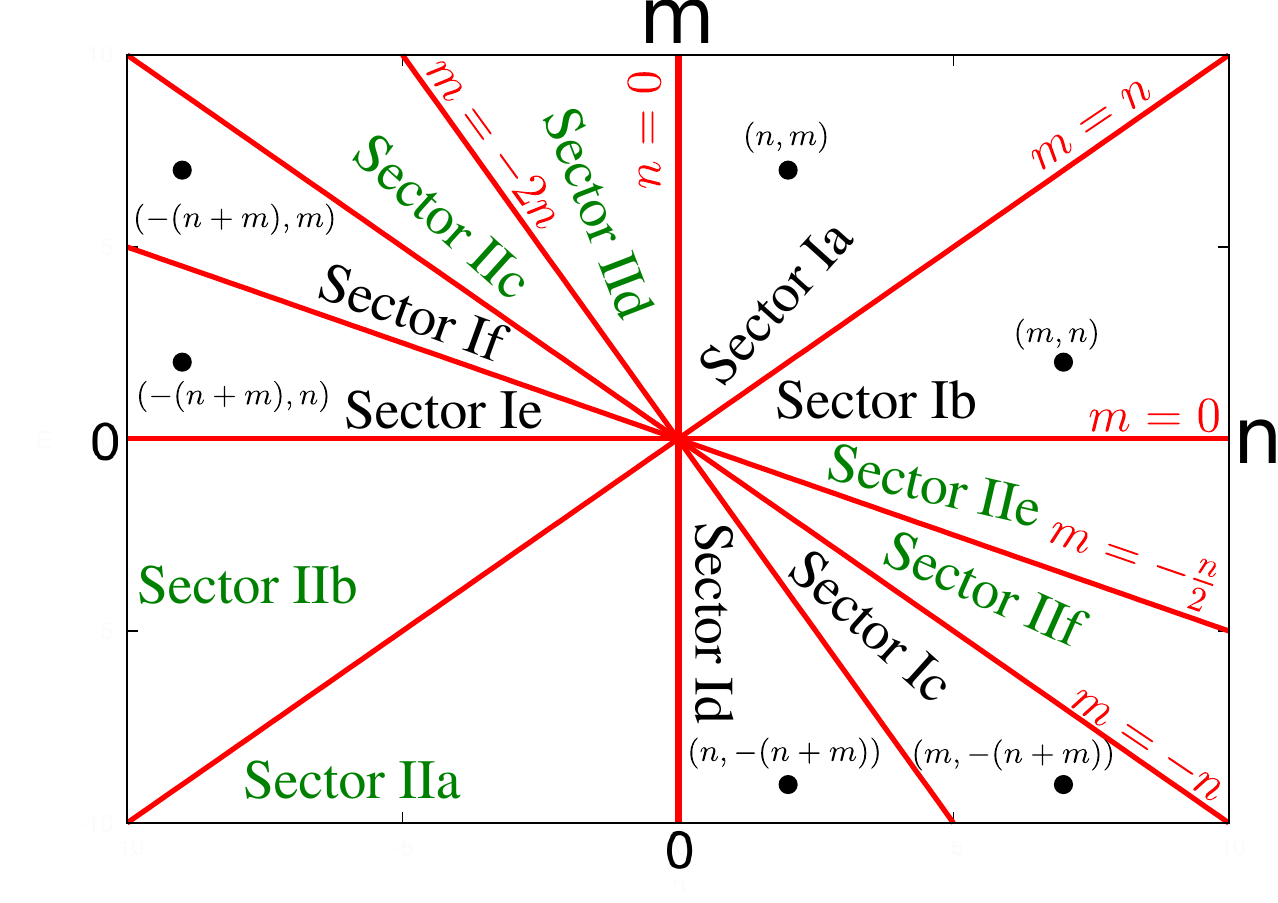}
\caption{Schematic diagram for explaining the derivation of Eq. (\ref{eq.appendix.fourier expansions of products of double sums in T3 square.06}).}
\label{fig.schematic for double summation of T3 square}
\end{center}
\end{figure}

This section is devoted to calculating certain terms in $\left(\mathcal{T}^{(3)}\right)^2$; specifically, the Fourier expansion of the product of the double summation terms appearing in Eqs. (\ref{eq.fourier expansion of T3.02}) and (\ref{eq.fourier expansion of T3.03}). Let us first consider products of the form
\begin{equation}
P=
\left(
\sum_{\substack{n=-\infty\\(\ne 0)}}^{\infty}
\sum_{\substack{m=-\infty\\(\ne 0,-n)}}^{\infty}
c_{-(n+m)}
c_{n}
c_{m}
X_{n,m}
\right)
\cdot
\left(
\sum_{\substack{n=-\infty\\(\ne 0)}}^{\infty}
\sum_{\substack{m=-\infty\\(\ne 0,-n)}}^{\infty}
c_{-(n+m)}
c_{n}
c_{m}
Y_{n,m}
\right)
.
\label{eq.appendix.fourier expansions of products of double sums in T3 square.01}
\end{equation}
Expanding $P$ out and keeping only the terms that do not vanish upon integration by the gaussian in Eq. (\ref{eq.IV.B.02}), we have
\begin{align}
P
\stackrel{\mathrm{n.v.}}{=}
&
\sum_{\substack{n=-\infty\\(\ne0) }}^{\infty} 
c_{-2n}c_{n}c_{n}X_{n,n}
\left[
c_{2n}c_{-n}c_{-n}Y_{-n,-n}
+
c_{-n}c_{-n}c_{2n}Y_{-n,2n}
+
c_{-n}c_{2n}c_{-n}Y_{2n,-n}
\right]
\,\,\,
+
\nonumber\\
&
\sum_{\substack{n=-\infty\\(\ne0) }}^{\infty} 
c_{n}c_{n}c_{-2n}X_{n,-2n}
\left[
c_{-n}c_{-n}c_{2n}Y_{-n,2n}
+
c_{2n}c_{-n}c_{-n}Y_{-n,-n}
+
c_{-n}c_{2n}c_{-n}Y_{2n,-n}
\right]
\,\,\,
+
\nonumber\\
&
\sum_{\substack{n=-\infty\\(\ne0) }}^{\infty} 
c_{n}c_{-2n}c_{n}X_{-2n,n}
\left[
c_{-n}c_{-n}c_{2n}Y_{-n,2n}
+
c_{2n}c_{-n}c_{-n}Y_{-n,-n}
+
c_{-n}c_{2n}c_{-n}Y_{2n,-n}
\right]
\,\,\,
+
\nonumber\\
&
\sum_{\substack{n=-\infty\\(\ne0) }}^{\infty} 
\sum_{\substack{m=-\infty\\(\ne0,n,-n,-\frac{n}{2},-2n) }}^{\infty} 
c_{-(n+m)}c_{n}c_{m}X_{n,m}
\left[
c_{n+m}c_{-n}c_{-m}Y_{-n,-m}
+
c_{m+n}c_{-m}c_{-n}Y_{-m,-n}
\right.
\nonumber\\
&
\left.
+
c_{-n}c_{n+m}c_{-m}Y_{(n+m),-m}
+
c_{-m}c_{n+m}c_{-n}Y_{(n+m),-n}
+
c_{-m}c_{-n}c_{n+m}Y_{-n,(n+m)}
+
c_{-n}c_{-m}c_{n+m}Y_{-m,(n+m)}
\right]
\label{eq.appendix.fourier expansions of products of double sums in T3 square.02}
\\
=
&
\,
\frac{1}{4}
\sum_{\substack{n=-\infty\\(\ne0) }}^{\infty}
c_{2n}c_{-2n}(c_{n}c_{-n})^2
\,
X_{-n,-n}^{\dagger}
Y_{n,n}^{\dagger}
+
\sum_{\substack{n=-\infty\\(\ne0) }}^{\infty} 
\sum_{\substack{m=-\infty\\(\ne0,n,-n,-\frac{n}{2},-2n) }}^{\infty} 
c_{n+m}c_{-(n+m)}c_{n}c_{-n}c_{m}c_{-m}
X_{n,m}Y^{\dagger}_{n,m}
\label{eq.appendix.fourier expansions of products of double sums in T3 square.03},
\end{align}
where
\begin{equation}
Y^{\dagger}_{n,m}
=
Y_{-n,-m} + Y_{-m,-n} + Y_{(n+m),-m} +
Y_{(n+m),-n} + Y_{-n,(n+m)} + Y_{-m,(n+m)}
.
\label{eq.appendix.fourier expansions of products of double sums in T3 square.04}
\end{equation}
Fig. \ref{fig.schematic for double summation of T3 square} is a schematic diagram showing the domain of the double summation in Eq. (\ref{eq.appendix.fourier expansions of products of double sums in T3 square.03}). One sums over all tuples of integers $(n,m)$ in the $n$-$m$ plane except those lying on the lines $n=0,m=0,m=n,m=-n,m=-2n$, and $m=-\frac{n}{2}$ (indicated by red solid lines). The summations for $m=n$, $m=-2n$, and $m=-\frac{n}{2}$ have already been separately accounted for by the first three lines of Eq. (\ref{eq.appendix.fourier expansions of products of double sums in T3 square.02}). 

We now simplify the double summation in Eq. (\ref{eq.appendix.fourier expansions of products of double sums in T3 square.03}). Notice that when $n,m>0$ and $n<m$, the terms $X_{n,m}$, $X_{m,n}$, $X_{m,-(n+m)}$, $X_{n,-(n+m)}$, $X_{-(n+m),n}$, and $X_{-(n+m),m}$ (from Sector Ia to Sector If, respectively) are all multiplied by the same term $c_{n+m}c_{-(n+m)}c_{n}c_{-n}c_{m}c_{-m}Y_{n,m}^{\dagger}$. Hence, we have
\begin{align}
&
\sum_{\mathrm{Sectors\,Ia\,to\,If}}
c_{n+m}c_{-(n+m)}c_{n}c_{-n}c_{m}c_{-m}X_{n,m}Y_{n,m}^{\dagger}
\nonumber\\
&
=
\sum_{\mathrm{Sector\,Ia}}
c_{n+m}c_{-(n+m)}c_{n}c_{-n}c_{m}c_{-m}
[
X_{n,m}+X_{m,n}+X_{m,-(n+m)}+X_{n,-(n+m)}+X_{-(n+m),n}+X_{-(n+m),m}
]
Y_{n,m}^{\dagger}
\nonumber\\
&
=
\frac{1}{2}
\sum_{n=1}^{\infty}
\sum_{\substack{m=1\\(\ne n)}}^{\infty}
c_{n+m}c_{-(n+m)}c_{n}c_{-n}c_{m}c_{-m}
X_{-n,-m}^{\dagger}
Y_{n,m}^{\dagger}
,
\label{eq.appendix.fourier expansions of products of double sums in T3 square.05}
\end{align}
Repeating the same procedure to Sectors IIa to IIf (simply let $(n,m)\rightarrow (-n,-m)$), Eq. (\ref{eq.appendix.fourier expansions of products of double sums in T3 square.03}) becomes
\begin{equation}
P
\stackrel{\mathrm{n.v.}}{=}
\frac{1}{4}
\sum_{\substack{n=-\infty\\(\ne0) }}^{\infty}
c_{2n}c_{-2n}(c_{n}c_{-n})^2
X_{-n,-n}^{\dagger}
Y_{n,n}^{\dagger}
+
\frac{1}{2}
\sum_{n=1}^{\infty}
\sum_{\substack{m=1\\(\ne n) }}^{\infty}
c_{n+m}c_{-(n+m)}c_{n}c_{-n}c_{m}c_{-m}
[
X^{\dagger}_{-n,-m}Y^{\dagger}_{n,m}
+
X^{\dagger}_{n,m}Y^{\dagger}_{-n,-m}
]
.
\label{eq.appendix.fourier expansions of products of double sums in T3 square.06}
\end{equation}

We now apply Eq. (\ref{eq.appendix.fourier expansions of products of double sums in T3 square.06}) to terms in $\left(\mathcal{T}^{(3)}\right)^2$ involving product of double summations. 

The first is the product of the double summation in Eq. (\ref{eq.fourier expansion of T3.02}) with itself. In this case, $X_{n,m}=Y_{n,m}=-\frac{1}{(2\pi)^2 m(n+m)}$. A straightforward calculation yields $Y_{n,n}^{\dagger}=Y_{n,m}^{\dagger}=0$, from which $X_{-n,-n}^{\dagger}=X_{-n,-m}^{\dagger}=X_{n,m}^{\dagger}=Y_{-n,-m}^{\dagger}=0$ immediately follows. Hence, 
\begin{equation}
\left(
-
\sum_{\substack{n=-\infty\\(\ne 0)}}^{\infty}
\sum_{\substack{m=-\infty\\(\ne 0,-n)}}^{\infty}
\frac{c_{-(n+m)}c_{n}c_{m}}{(2\pi)^2m(n+m)}
\right)^2
\stackrel{\mathrm{n.v.}}{=}
0
.
\label{eq.appendix.fourier expansions of products of double sums in T3 square.07}
\end{equation}

The second is the product of the double summation in Eq. (\ref{eq.fourier expansion of T3.02}) with the double summation in Eq. (\ref{eq.fourier expansion of T3.03}). In this case, $X_{n,m}=-\frac{1}{(2\pi)^2 m(n+m)}$ and $Y_{n,m}=\lambda_{nm}$. From $X_{-n,-n}^{\dagger}=X_{-n,-m}^{\dagger}=X_{n,m}^{\dagger}=0$, we have
\begin{equation}
\left(
-
\sum_{\substack{n=-\infty\\(\ne 0)}}^{\infty}
\sum_{\substack{m=-\infty\\(\ne 0,-n)}}^{\infty}
\frac{c_{-(n+m)}c_{n}c_{m}}{(2\pi)^2m(n+m)}
\right)
\left(
\sum_{\substack{n=-\infty\\(\ne 0)}}^{\infty}
\sum_{\substack{m=-\infty\\(\ne 0,-n)}}^{\infty}
c_{-(n+m)}c_{n}c_{m}
\lambda_{nm}
\right)
\stackrel{\mathrm{n.v.}}{=}
0
.
\label{eq.appendix.fourier expansions of products of double sums in T3 square.08}
\end{equation}

The third is the product of the double summation in Eq. (\ref{eq.fourier expansion of T3.03}) with itself. In this case, $X_{n,m}=Y_{n,m}=\lambda_{nm}$. We simply state the result.
\begin{align}
&
\left(
\sum_{\substack{n=-\infty\\(\ne 0)}}^{\infty}
\sum_{\substack{m=-\infty\\(\ne 0,-n)}}^{\infty}
c_{-(n+m)}c_{n}c_{m}
\lambda_{nm}
\right)^2
\nonumber\\
&
\stackrel{\mathrm{n.v.}}{=}
\sum_{\substack{n=-\infty \\ (\ne 0) }}^{\infty}
c_{2n}c_{-2n}(c_{n}c_{-n})^2
\left[
\frac{12(2\varepsilon)^2\sinh\varepsilon}{[n][2n]}
\right]^2
\nonumber\\
&
+
\sum_{n=1}^{\infty}
\sum_{\substack{m=1 \\ (\ne n)}}^{\infty}
c_{n+m}c_{-(n+m)}c_nc_{-n}c_mc_{-m}
\left[
\frac{8(2\varepsilon)^2\sinh\varepsilon}{[n+m]}
\left(
\frac{1}{[n]}
+
\frac{1}{[m]}
+
\frac{(2\pi n)(2\pi m)+(2\varepsilon)^2}{[n][m]}
\right)
\right]
^2
.
\label{eq.appendix.fourier expansions of products of double sums in T3 square.09}
\end{align}

\section{Validity of Eq. (\ref{eq.V.A.06}) for the model Eq. (\ref{eq.II.A.01}) in the paramagnetic phase}
\label{sec.appendix.ZQ in para phase}

\subsection{When $J=0$}
\label{sec.appendix.ZQ in para phase. J=0}

The ground-state of $H$ is
\begin{equation}
|E_0\rangle_{J=0}
=\prod_{i=1}^{N}
|\sigma_i^x=+1\rangle,
\label{eq.ZQ para phase.01}
\end{equation}
i.e., a direct-product state where all spins point along the positive $x$-direction. $|E_0\rangle_{J=0}$ is non-degenerate and has parity $+1$. The first excited-state is $N$-fold degenerate. Let $|i\rangle$ denote the state where the $i$th spin in $|E_0\rangle_{J=0}$ is flipped. The level is spanned by the set $\{|i\rangle\}$. Any linear combination of the $|i\rangle$'s has parity $-1$. Hence, condition 2 is satisfied when $J=0$.

\subsection{When $J$ is turned on}
\label{sec.appendix.ZQ in para phase. J not 0}

Since $Q$ is a conserved quantity, the parity of $|E_0\rangle_{J=0}$ cannot change when $J$ is turned on. Its non-degeneracy is also maintained until the phase transition point where it collides with the first excited-state.

For the first excited-state, first-order perturbation theory shows that the subspace that splits away from the original $N$-dimensional one when $J$ is turned on is
\begin{equation}
|E_1\rangle_{J=0}
=
\sum_{i=1}^N |i\rangle.
\label{eq.ZQ para phase.02}
\end{equation}
$|E_1\rangle_{J=0}$ is non-degenerate and has parity $-1$. Its parity must also be conserved as $J$ increases, until collision with $|E_0\rangle$ at the transition point. Hence, condition 2 is satisfied when $J$ is turned on in the paramagnetic regime.

To conclude, Eq. (\ref{eq.V.A.06}) is valid for the model Eq. (\ref{eq.II.A.01}) in the paramagnetic phase.

\section{Validity of using $A_z$ and $A_y$ in Eq. (\ref{eq.VI.A.02}) for the model Eq. (\ref{eq.II.A.01})}
\label{sec.appendix.ZA}

\subsection{Concerning $A_z$ and $A_y$ satisfying the conditions of selection rule 1 in the paramagnetic phase}
\label{sec.appendix.for Az and Ay}

Condition 1 has been shown in Appendix \ref{sec.appendix.ZQ in para phase} to be satisfied in the paramagnetic phase. It is straightforward to verify that $A_z$ and $A_y$ anti-commute with $Q$. It remains to check conditions 3 in the paramagnetic phase.

When $J=0$, with the notations of Appendix \ref{sec.appendix.ZQ in para phase}, one has
\begin{equation}
\sum_{i=1}^{N} |_{J=0}\langle E_0|A_{\mu}|i\rangle|^2=1,
\label{eq.concerning ZA.01}
\end{equation}
for $\mu=z$ and $y$, which is non-zero. When $J$ is turned on, the ground-state $|E_0\rangle$ and the first excited-state $|E_1\rangle$ are both non-degenerate and they lie in the sector with total angular momentum $N/2$. We diagonalized the Hamiltonian Eq. (\ref{eq.II.A.01}) in this sector and study the matrix elements numerically. Figs. \ref{fig.concerning nonvanishing of matrix elements}(a) and (b) show the absolute values of $\langle E_0|A_z|E_1\rangle$ and $\langle E_0|A_y|E_1\rangle$, respectively, for $N=10, 100$, and 1000. We see that the matrix elements are non-zero in the paramagnetic regime $\Gamma>2J$.

\subsection{Concerning $A_y$ satisfying the conditions of selection rule 2 in both phases}
\label{sec.appendix.for Ay only}

Condition 1 of selection rule 2 is satisfied by $A_y$, since $[H,A_z]=2i\Gamma A_y$. Concerning condition 2, as the behavior of $\sum_{a,b}|\langle E_0^a|A| E_1^b\rangle|^2$ in the paramagnetic phase has already been discussed in the previous section, we now discuss the ferromagnetic phase. 

When $\Gamma=0$, the doubly-degenerate ground-state is spanned by
\begin{equation}
|E_0^{\pm}\rangle_{\Gamma=0}
=
\frac{1}{\sqrt{2}}
\left(
\prod_{i=1}^N
|\sigma_i^z=+1\rangle
\pm
\prod_{i=1}^N
|\sigma_i^z=-1\rangle
\right),
\label{eq.concerning ZA.02}
\end{equation}
where the superscript $\pm$ labels the parity quantum number. The first excited-state is $2N$-fold degenerate. Let $|i\rangle_{\pm}$ denote the state where the $i$th spin in $|E_0^{\pm}\rangle_{\Gamma=0}$ is flipped (e.g., $\sigma_i^y|E_0^{\pm}\rangle_{\Gamma=0}$). The level is spanned by the set $\{|i\rangle_{\pm}\}$. Hence, the required $\sum_{a,b}|\langle E_0^a|A| E_1^b\rangle|^2$ becomes
\begin{equation}
\sum_{p=\pm}
\sum_{i=1}^N
\sum_{p^{\prime}=\pm}
|_{\Gamma=0}
\langle 
E_0^p
|A_y|
i
\rangle_{p^{\prime}}
|^2
=2.
\label{eq.concerning ZA.03}
\end{equation}
which is non-zero. 

When $\Gamma$ is turned on, second-order perturbation theory shows that the subspace that splits away to form the first excited-state is spanned by
\begin{equation}
|E_1^{\pm}\rangle_{\Gamma=0}
=
\frac{1}{\sqrt{2N}}
\left(
\sum_{i=1}^N	
|i\rangle_+
\pm
|i\rangle_-
\right).
\label{eq.concerning ZA.04}
\end{equation}
Hence, in the ferromagnetic phase the ground and first excited-states are both doubly-degenerate. The sum $\sum_{a,b}|\langle E_0^a|A| E_1^b\rangle|^2$ therefore consists of four matrix elements, $\langle E_0^+ |A_y|E_1^+ \rangle$, $\langle E_0^- |A_y|E_1^- \rangle$, $\langle E_0^- |A_y|E_1^+ \rangle$, and $\langle E_0^+ |A_y|E_1^- \rangle$. The first two vanish because of selection rule 1. It remains to check, numerically, that the latter two are non-zero.

The Hamiltonian Eq. (\ref{eq.II.A.01}) is again diagonalized in the sector with total angular momentum $N/2$, and the eigenvectors of the four lowest energies are used to diagonalize the parity operator $Q$ to obtain the parity eigenvectors. Figs. \ref{fig.concerning nonvanishing of matrix elements}(c) and (d) show the absolute values of $\langle E_0^- |A_y|E_1^+\rangle$ and $\langle E_0^+ |A_y|E_1^-\rangle$, respectively, for $N=10, 100,$ and 1000. It is seen that they are non-zero in the ferromagnetic regime $\Gamma<2J$.

\section{Calculation of first-order approximation of $\mathcal{T}_{2y}$}
\label{sec.appendix.calculating T2y}

We first consider static approximation. The summand corresponding to $\sigma=+1$ in Eq. (\ref{eq.VI.C.06}) is
\begin{equation}
\left(
\begin{array}{cc}
  a_+^{(0)}  &  a_-^{(0)}  \\
\end{array}
\right)
\left(
\begin{array}{cc}
  e^{2\varepsilon}  & 0 \\
  0  & e^{-2\varepsilon} \\
\end{array}
\right)
\left(
\begin{array}{cc}
  0  & -i \\
  i  &  0 \\
\end{array}
\right)
\left(
\begin{array}{cc}
  e^{2\varepsilon}  & 0 \\
  0  & e^{-2\varepsilon} \\
\end{array}
\right)
\left(
\begin{array}{c}
 a_+^{(0)} \\
 a_-^{(0)} \\
\end{array}
\right).
\label{eq.calculate T2y.01}
\end{equation}
The above and the summand corresponding to $\sigma=-1$ are both identically zero. Adding, we have $\mathcal{T}_{2y}\stackrel{\mathrm{s.a.}}{\longrightarrow}0$.

We now calculate the $\lambda^1$ term. At $t=1$, we have
\begin{equation}
|+1(1)\rangle
\stackrel{\mathrm{1st}}{\longrightarrow}
\left(
\begin{array}{c}
 a_+^{(0)} \,e^{2\varepsilon} \\
 a_-^{(0)} e^{-2\varepsilon} \\
\end{array}
\right)
+
\lambda
(4\beta J)
\left(
\begin{array}{c}
 (\alpha a_+^{(0)} \int_0^1 m_d(t) dt  +  \gamma a_-^{(0)} \int_0^1 m_d(t) e^{-4\varepsilon t} dt)\, e^{2\varepsilon}\\
 (\gamma a_+^{(0)} \int_0^1 m_d(t) e^{4\varepsilon t} dt -\alpha a_-^{(0)} \int_0^1 m_d(t) dt)\, e^{-2\varepsilon} \\
\end{array}
\right)
.
\label{eq.calculate T2y.02}
\end{equation}
Multiplying by $\sigma^y$, we have
\begin{equation}
\sigma^y
|+1(1)\rangle
\stackrel{\mathrm{1st}}{\longrightarrow}
\left(
\begin{array}{c}
-i a_-^{(0)} e^{-2\varepsilon} \\
i a_+^{(0)} e^{2\varepsilon} \\
\end{array}
\right)
+
i
\lambda
(4\beta J)
\left(
\begin{array}{c}
 (\alpha a_-^{(0)} \int_0^1 m_d(t) dt-\gamma a_+^{(0)} \int_0^1 m_d(t) e^{4\varepsilon t} dt)\, e^{-2\varepsilon} \\
 (\alpha a_+^{(0)} \int_0^1 m_d(t) dt  +  \gamma a_-^{(0)} \int_0^1 m_d(t) e^{-4\varepsilon t} dt)\, e^{2\varepsilon}\\
\end{array}
\right)
.
\label{eq.calculate T2y.03}
\end{equation}
From here, to get the $\lambda^1$ term, we first propagate the second term of Eq. (\ref{eq.calculate T2y.03}) at static approximation to $t=2$ and take the inner product,
\begin{equation}
i
\lambda
(4\beta J)
\left(
\begin{array}{cc}
  a_+^{(0)}  &  a_-^{(0)}  \\
\end{array}
\right)
\left(
\begin{array}{cc}
  e^{2\varepsilon}  & 0 \\
  0  & e^{-2\varepsilon} \\
\end{array}
\right)
\left(
\begin{array}{c}
 (\alpha a_-^{(0)} \int_0^1 m_d(t) dt-\gamma a_+^{(0)} \int_0^1 m_d(t) e^{4\varepsilon t} dt)\, e^{-2\varepsilon} \\
 (\alpha a_+^{(0)} \int_0^1 m_d(t) dt  +  \gamma a_-^{(0)} \int_0^1 m_d(t) e^{-4\varepsilon t} dt)\, e^{2\varepsilon}\\
\end{array}
\right)
.
\label{eq.calculate T2y.04}
\end{equation}
Summing with the contribution from the $\sigma=-1$ term and using Eqs. (\ref{eq.III.C.06}) and (\ref{eq.III.C.07}), we obtain the first and second terms inside the parenthesis of Eq. (\ref{eq.VI.D.03}). To obtain the third and fourth terms inside the parenthesis, propagate the first term of Eq. (\ref{eq.calculate T2y.03}) at first order to $t=2$, 
\begin{equation}
\lambda
(4\beta J)
\left(
\begin{array}{cc}
 e^{2\varepsilon}   &  0  \\
 0   &  e^{-2\varepsilon} \\
\end{array}
\right)
\left(
\begin{array}{cc}
 \alpha \int_0^1 m_d(1+\tau) d\tau  &  \gamma \int_0^1 m_d(1+\tau) e^{-4\varepsilon \tau} d\tau \\
 \gamma \int_0^1 m_d(1+\tau) e^{4\varepsilon \tau}d\tau   &  - \alpha \int_0^1 m_d(1+\tau) d\tau\\
\end{array}
\right)
\left(
\begin{array}{c}
-i a_-^{(0)} e^{-2\varepsilon} \\
i a_+^{(0)} e^{2\varepsilon} \\
\end{array}
\right)
,
\label{eq.calculate T2y.05}
\end{equation}
where we have translated the time variable in $m_d(t)$ forward by 1 because in Eq. (\ref{eq.III.C.08}) the lower integration limit needs to start from zero. Changing the integration variable back to $t$ via $t=1+\tau$, we have $\int_0^1 m_d(1+\tau) d\tau=\int_1^2 m_d(t) dt$ and $\int_0^1 m_d(1+\tau) e^{\pm4\varepsilon \tau} d\tau=\int_1^2 m_d(t) e^{\pm4\varepsilon (t-1)} dt$. Taking the inner product, adding with the contribution from $\sigma=-1$, and using Eqs. (\ref{eq.III.C.06}) and (\ref{eq.III.C.07}), we obtain the third and fourth terms.

\section{Calculation of $Z_{A_z}$ in the paramagnetic phase}
\label{sec.appendix.calculation of ZAz}

In the paramagnetic phase, $m_s=0$, so $\gamma=-1$ and $\alpha=0$. Inserting the expansion Eq. (\ref{eq.VI.E.01}) into Eqs. (\ref{eq.VI.D.05}) and (\ref{eq.VI.D.06}), we have
\begin{equation}
\frac{\mathcal{T}_0 \mathcal{T}_{3z} + (N-1)\mathcal{T}_{1z} \mathcal{T}_{2z} }{(\mathcal{T}_0)^2}
=
\mathrm{sech}4\varepsilon
+
(g')^2
\sum_{n=-\infty}^{\infty}
c_n c_{-n}
\frac{(-1)^n }{[(\pi n)^2+(4\varepsilon)^2]^2}
+
O(N^{-1/2})
,
\label{eq.calculation of ZAz.01}
\end{equation}
where once again the cross terms have been dropped. Inserting Eq. (\ref{eq.calculation of ZAz.01}) into Eq. (\ref{eq.VI.E.02}), and following the same derivation as that for $Z_{A_y}$, we obtain
\begin{equation}
Z_{A_z}
\stackrel{O(1)}{\longrightarrow}
e^{-\beta N f'_s}
\,
\frac{\sinh4\varepsilon \tanh4\varepsilon}{\sinh^2 4\sqrt{\varepsilon^2-\frac{g'}{16}}}
\,
\frac{\varepsilon}{\sqrt{\varepsilon^2-\frac{g'}{16}}}
.
\label{eq.calculation of ZAz.02}
\end{equation}
Like $Z_{A_y}$, Eq. (\ref{eq.calculation of ZAz.02}) is valid everywhere except at the critical point where it diverges due to the vanishing of $\sqrt{\varepsilon^2-\frac{g'}{16}}$. The result Eq. (\ref{eq.calculation of ZAz.02}) is the same as that for $Z_{A_y}$ except for the last factor $\frac{\varepsilon}{\sqrt{\varepsilon^2-\frac{g'}{16}}}$ which goes to zero when taking the limit $\beta\rightarrow\infty$ in Eq. (\ref{eq.VI.A.02}). Hence, we once again obtain Eq. (\ref{eq.VI.F.02}) for the energy gap, this time only in the paramagnetic phase.

\begin{thebibliography}{99} 


\bibitem{Perdomo-Ortiz12} A. Perdomo-Ortiz, N. Dickson, M. Drew-Brook, G. Rose, and A. Aspuru-Guzik, Scientific Reports \textbf{2}, 571 (2012). 

\bibitem{Liu15} C. W. Liu, A. Polkovnikov, and A. W. Sandvik, Phys. Rev. Lett. \textbf{114}, 147203 (2015).

\bibitem{Nishimura16} K. Nishimura, H. Nishimori, A. J. Ochoa, and H. G. Katzgraber, Phys. Rev. E \textbf{94}, 032105 (2016).

\bibitem{Takahashi17} K. Takahashi, Phys. Rev. A \textbf{95}, 012309 (2017).


\bibitem{Kadowaki98} T. Kadowaki and H. Nishimori, Phys. Rev. E \textbf{58}, 5355 (1998).

\bibitem{Farhi01} E. Farhi, J. Goldstone, S. Gutmann, J. Lapan, A. Lundgren, and D. Preda, Science \textbf{292}, 472 (2001).

\bibitem{Das08} A. Das and B. K. Chakrabarti, Rev. Mod. Phys. \textbf{80}, 1061 (2008).


\bibitem{Jorg08} T. J\"{o}rg, F. Krzakala, J. Kurchan, and A.C. Maggs,  Phys. Rev. Lett. \textbf{101}, 147204 (2008). 

\bibitem{Jorg10} T. J\"{o}rg, F. Krzakala, J. Kurchan, A.C. Maggs, and J. Pujos,  EuroPhys. Lett. \textbf{89}, 40004 (2010).

\bibitem{Bapst12} V. Bapst and G. Semerjian, J. Stat. Mech., P06007 (2012).

\bibitem{Seki12} Y. Seki and H. Nishimori, Phys. Rev. E \textbf{85}, 051112 (2012).

\bibitem{Bray80} A.J. Bray and M.A. Moore, J. Phys. C: Solid St. Phys. \textbf{13}, L655 (1980).

\bibitem{Thirumalai89} D. Thirumalai, Q. Li, and T.R. Kirkpatrick, J. Phys. A: Math. Theor. \textbf{22}, 3339 (1989).

\bibitem{Nishimori96} H. Nishimori and Y. Nonomura, J. Phys. Soc. Jpn. \textbf{65}, 3780 (1996). 

\bibitem{Dobrosavljevic90} V. Dobrosavljevi\'{c} and D. Thirumalai, J. Phys. A: Math. Theor. \textbf{23}, L767 (1990).

\bibitem{Goldschmidt90a} Y.Y. Goldschmidt, Phys. Rev. B \textbf{41}, 4858 (1990).

\bibitem{DeCesare96} L. De Cesare, K. Lukierska-Walasek, I. Rabuffo, and K. Walasek, J. Phys. A: Math. Theor. \textbf{29}, 1605 (1996).

\bibitem{Obuchi07} T. Obuchi, H. Nishimori, and D. Sherrington, J. Phys. Soc. Jpn. \textbf{76} 054002 (2007).

\bibitem{Takahashi07} K. Takahashi, Phys. Rev. B \textbf{76}, 184422 (2007).

\bibitem{Usadel87} K. D. Usadel and B. Schmidtz, Solid State Commun. \textbf{64}, 975 (1987).

\bibitem{Goldschmidt90b} Y. Y. Goldschmidt and P. Y. Lai, Phys. Rev. Lett. \textbf{64}, 2467 (1990). 

\bibitem{Miller93} J. Miller and D. A. Huse,  Phys. Rev. Lett. \textbf{70}, 3147 (1993). 

\bibitem{Grempel98} D. R. Grempel and M. J. Rozenberg,  Phys. Rev. Lett. \textbf{80}, 389 (1998). 

\bibitem{Rozenberg98} M. J. Rozenberg and D. R. Grempel, Phys. Rev. Lett. \textbf{81}, 2550 (1998).

\bibitem{Sei07} S. Suzuki, H. Nishimori, and M. Suzuki, Phys. Rev. E \textbf{75}, 051112 (2007).

\bibitem{Ohzeki11} M. Ohzeki and H. Nishimori, J. Comp. Theor. Nanoscience \textbf{8}, 963 (2011).

\bibitem{Seoane12} B. Seoane and H. Nishimori, J. Phys. A: Math. Theor. \textbf{45}, 435301 (2012).

\bibitem{Koh16} Y. W. Koh, Phys. Rev. B \textbf{93}, 134202 (2016).

\bibitem{Dusuel04} S. Dusuel and J. Vidal, Phys. Rev. Lett. \textbf{93}, 237204 (2004).

\bibitem{Dusuel05} S. Dusuel and J. Vidal, Phys. Rev. B \textbf{71}, 224420 (2005).

\bibitem{Wegner94} F. Wegner, Ann. Phys. \textbf{3}, 77 (1994).

\bibitem{Glazek93} S. D. Glazek and K. G. Wilson, Phys. Rev. D \textbf{48}, 5863 (1993).

\bibitem{Glazek94} S. D. Glazek and K. G. Wilson, Phys. Rev. D \textbf{49}, 4214 (1994).

\bibitem{note. why sqrt of N} See the discussion in Sec. \ref{sec.sub.E0 2nd order} for the origin of the square root.

\bibitem{Takahashi04} S. Takahashi and K. Takatsuka, Phys. Rev. A \textbf{69}, 022110 (2004).

\bibitem{Goldstein01} H. Goldstein, C. Poole, and J. Safko, \emph{Classical Mechanics}, 3rd ed. (Pearson, New York, 2001).

\bibitem{Brewer97} M. L. Brewer, J. S. Hulme, and D. E. Manolopoulos, J. Chem. Phys. \textbf{106}, 4832 (1997).

\bibitem{Schulman96} L. S. Schulman, \emph{Techniques and Applications of Path Integration} (Wiley-Interscience, New York, 1996).

\bibitem{Bandyopadhyay08} S. Bandyopadhyay and M. Cahay, \emph{Introduction to Spintronics} (CRC Press, Boca Raton, 2008).

\bibitem{Ballentine98} L. E. Ballentine, \emph{Quantum Mechanics: A Modern Development} (World Scientific, Singapore, 1998). See p. 349 for a discussion on time-dependent perturbation theory. 

\bibitem{note.concerning generalized partition function} Eq. (\ref{eq.intro.02}) is not completely general because one may need to consider `cross-terms' such as $e^{-\beta E_0}\times e^{-\beta E_1}$ on the right hand side of the equation. However, the overall argument remains the same and Eq. (\ref{eq.intro.02}) is sufficient simply for conveying the main idea.  

\bibitem{Zinn-Justin05} J. Zinn-Justin, \emph{Path Integrals in Quantum Mechanics} (Oxford University Press, New York, 2005). See p. 229 for a discussion on level splitting. 

\bibitem{Hogg03} T. Hogg, Phys. Rev. A \textbf{67}, 022314 (2003).

\bibitem{Takahashi10a} K. Takahashi and Y. Matsuda, J. Phys. Soc. Jpn. \textbf{79}, 043712 (2010). 

\bibitem{Takahashi10b} K. Takahashi and Y. Matsuda, J. Phys. : Conference Series \textbf{233}, 012008 (2010).

\bibitem{Young08} A.P. Young, S. Knysh, and V.N. Smelyanskiy, Phys. Rev. Lett. \textbf{101}, 170503 (2008).

\bibitem{Young10} A.P. Young, S. Knysh, and V.N. Smelyanskiy, Phys. Rev. Lett. \textbf{104}, 020502 (2010).

\bibitem{Das06} A. Das, K. Sengupta, D. Sen, and B.K. Chakrabarti, Phys. Rev. B \textbf{74}, 144423 (2006).

\bibitem{note.notation for measure} The discrete form of $\int Dm_d(t)$ is $\int \prod_{\kappa} \lambda\sqrt{\frac{\beta J N}{\pi M}} d(m_d)_{\kappa}$. The latter is more useful when calculating the jacobian associated with the change of variables to Fourier coefficients. 


\bibitem{note.concerning sech2e term} From the Fourier expansion Eq. (\ref{eq.IV.A.03}), $M_0=c_0$ and $M_{00}=\frac{c^2_0}{2}$, and $M_{00}-\frac{1}{2}\tanh^2\varepsilon(M_0)^2=\frac{c_0^2}{2}\mathrm{sech}^2\varepsilon$.

\bibitem{note.complex gaussian integral formula} The gaussian integral takes the form $\int dcdc^* e^{-g cc^*}=\frac{\pi}{g}$.

\bibitem{note.reason for dropping V3/sqrtN term} In Eq. (\ref{eq.IV.B.02}), we have dropped the term $\frac{1}{\sqrt{N}}V_3$ since it contains odd powers of $c_n$ and hence vanishes when integrated over by the gaussian function.

\bibitem{note.instructions for L1 and L2} To recall, for $\mathcal{T}^{(1)}$, we have $M_0=c_0$; for $\mathcal{T}^{(2)}$, we have $M_{00}=\frac{c_0^2}{2}$ and Eq. (\ref{eq.IV.A.04}).

\bibitem{note.explaining origin of each of the terms making up N-1 term} The term $z_4^1$ stems from the addition of the square of Eq. (\ref{eq.IV.A.04}) and the third and fourth lines of Eq. (\ref{eq.fourier expansion of T4.02}) involving $(c_nc_{-n})^2$ and $c_nc_{-n}c_mc_{-m}$. The term $z_4^2$ stems from Eq. (\ref{eq.fourier expansion of T4.08}). The term $z_3^2$ stems from the second term of Eq. (\ref{eq.appendix.fourier expansions of products of double sums in T3 square.09}). 

\bibitem{note.concerning computing double sums of z42 and z32 numerically} A double sum, denoted $\Xi$, is computed for several large values of $\beta$ while keeping all the other parameters fixed. Fitting a straight line to $\ln\Xi=-s_1\ln\beta+s_2$, we determined $s_1$ and $s_2$. This gives $\Xi=e^{s_2}\cdot\beta^{-s_1}$, the asymptotic form of $\Xi$ as $\beta\rightarrow\infty$. The term $\beta^{-s_1}$ will ultimately be cancelled by other $\beta$'s coming from the prefactor, leaving $e^{s_2}$ as the contribution to $z_4^2$ or $z_3^2$.

\bibitem{note.form of sigmax in energy basis} Note that $\sigma^{x}$ takes the form ${-\gamma\,\,\alpha\choose \alpha \,\,\,\, \gamma}$ in the basis where $\mathcal{H}_s$ is diagonal.


\bibitem{note.proof of selection rule of A} J. J. Sakurai, \emph{Modern Quantum Mechanics}, revised ed. (Addison-Wesley, Reading, 1994). The following proof is taken from p. 259. $\langle q|A|q^{\prime}\rangle=\langle q|Q^{-1}QAQ^{-1}Q|q^{\prime}\rangle=\langle q|q(-1)AQQ^{-1}q^{\prime}|q^{\prime}\rangle=-qq^{\prime}\langle q|A|q^{\prime}\rangle$. If $q=q^{\prime}$, then $qq^{\prime}=q^2=1$, so $\langle q|A|q\rangle=0$. Hence, $\langle q|A|q^{\prime}\rangle$ can be non-zero only if $q=-q^{\prime}$.

\bibitem{note.proof of selection rule using Aprime} Proof: From Eq. (\ref{eq.VI.A.04}), $\langle E_n^{a}|[H,A^{\prime}]|E_n^{b}\rangle=c\langle E_n^{a}|A|E_n^{b}\rangle$; but $\langle E_n^{a}|[H,A^{\prime}]|E_n^{b}\rangle=(E_n-E_n)\langle E_n^{a}|A^{\prime}|E_n^{b}\rangle=0$. Eq. (\ref{eq.VI.A.05}) follows.


\bibitem{note.explanation for scaling of two in J G and integration limits} The change in the upper integration limit is due to the two $e^{-2\beta H}$s. The change in $J$ and $\Gamma$ is due to the factor of 2 in the exponent of $e^{-2\beta H}$.

\bibitem{note. gaussian formula for second moment} Using $\int dcdc^* (cc^*) e^{-gcc^*}=\frac{\pi}{g^2}$.

\bibitem{note.concerning no collision between upper energy levels} We have assumed that the first excited-state does not collide with any other higher-energy levels before reaching the critical point; if that happens, the first excited-state may change parity, and the relation Eq. (\ref{eq.V.A.06}) will no longer hold. 

\bibitem{note.concerning vanishing of cncncncn term upon summation} Note that although the term $c_nc_nc_nc_n$ is not zero when integrated over  by the gaussian, calculations reveal that, as far as $\mathcal{T}^{(4)}$ is concerned, one obtains zero when summing over contributions coming from this term.

\end{thebibliography}
\end{document}